\documentclass[twocolumn]{aastex63}

\usepackage{float}
\usepackage{graphicx}
\usepackage{xcolor}
\usepackage{amsmath}
\usepackage{verbatim}

\newcommand\ms{$\textrm{m~s}^{-1}$}
\newcommand\cms{$\textrm{cm~s}^{-1}$}

\begin{document}

\title{A Harsh Test of Far-Field Scrambling with the Habitable Zone Planet Finder and the Hobby Eberly Telescope}

\author[0000-0001-8401-4300]{Shubham Kanodia}
\affiliation{Department of Astronomy \& Astrophysics, 525 Davey Laboratory, The Pennsylvania State University, University Park, PA, 16802, USA}
\affiliation{Center for Exoplanets and Habitable Worlds, 525 Davey Laboratory, The Pennsylvania State University, University Park, PA, 16802, USA}
\affiliation{Penn State Extraterrestrial Intelligence Center, 525 Davey Laboratory, The Pennsylvania State University, University Park, PA, 16802, USA}

\author[0000-0003-1312-9391]{Samuel Halverson}
\affiliation{Jet Propulsion Laboratory, California Institute of Technology, 4800 Oak Grove Drive, Pasadena, CA 91109, USA}

\author[0000-0001-6160-5888]{Joe P.\ Ninan}
\affiliation{Department of Astronomy \& Astrophysics, 525 Davey Laboratory, The Pennsylvania State University, University Park, PA, 16802, USA}
\affiliation{Center for Exoplanets and Habitable Worlds, 525 Davey Laboratory, The Pennsylvania State University, University Park, PA, 16802, USA}

\author[0000-0001-9596-7983]{Suvrath Mahadevan}
\affiliation{Department of Astronomy \& Astrophysics, 525 Davey Laboratory, The Pennsylvania State University, University Park, PA, 16802, USA}
\affiliation{Center for Exoplanets and Habitable Worlds, 525 Davey Laboratory, The Pennsylvania State University, University Park, PA, 16802, USA}

\author[0000-0001-7409-5688]{Gudmundur Stefansson}
\affiliation{Henry Norris Russell Fellow}
\affiliation{Department of Astrophysical Sciences, Princeton University, 4 Ivy Lane, Princeton, NJ 08540, USA}

\author[0000-0001-8127-5775]{Arpita Roy}
\affiliation{Space Telescope Science Institute, 3700 San Martin Dr, Baltimore, MD 21218, USA}

\author[0000-0002-4289-7958]{Lawrence W. Ramsey}
\affiliation{Department of Astronomy \& Astrophysics, 525 Davey Laboratory, The Pennsylvania State University, University Park, PA, 16802, USA}
\affiliation{Center for Exoplanets and Habitable Worlds, 525 Davey Laboratory, The Pennsylvania State University, University Park, PA, 16802, USA}

\author[0000-0003-4384-7220]{Chad F.\ Bender}
\affiliation{Steward Observatory, The University of Arizona, 933 N.\ Cherry Ave, Tucson, AZ 85721, USA}

\author[0000-0001-9165-8905]{Steven Janowiecki}
% \affiliation{McDonald Observatory and Department of Astronomy, The University of Texas at Austin}
\affiliation{University of Texas at Austin, McDonald Observatory, TX 79734, USA}

\author[0000-0001-9662-3496]{William D. Cochran}
\affiliation{McDonald Observatory and Department of Astronomy, The University of Texas at Austin}
\affiliation{Center for Planetary Systems Habitability, The University of Texas at Austin}

\author[0000-0002-2144-0764]{Scott A. Diddams}
\affiliation{Time and Frequency Division, National Institute of Standards and Technology, 325 Broadway, Boulder, CO 80305, USA}
\affiliation{Department of Physics, University of Colorado, 2000 Colorado Avenue, Boulder, CO 80309, USA}

\author{Niv Drory}
\affiliation{University of Texas at Austin, McDonald Observatory, TX 79734, USA}

\author[0000-0002-7714-6310]{Michael Endl}
\affiliation{McDonald Observatory and Department of Astronomy, The University of Texas at Austin}
\affiliation{Center for Planetary Systems Habitability, The University of Texas at Austin}

\author[0000-0001-6545-639X]{Eric B.\ Ford}
\affiliation{Department of Astronomy \& Astrophysics, 525 Davey Laboratory, The Pennsylvania State University, University Park, PA, 16802, USA}
\affiliation{Center for Exoplanets and Habitable Worlds, 525 Davey Laboratory, The Pennsylvania State University, University Park, PA, 16802, USA}
\affiliation{Institute for Computational and Data Sciences, The Pennsylvania State University, University Park, PA 16803, USA}
\affiliation{Center for Astrostatistics,  525 Davey Laboratory, The Pennsylvania State University, University Park, PA 16803, USA\\}
\affiliation{Institute for Advanced Study, 1 Einstein Dr, Princeton, NJ 08540}

\author[0000-0002-1664-3102]{Fred Hearty}
\affiliation{Department of Astronomy \& Astrophysics, 525 Davey Laboratory, The Pennsylvania State University, University Park, PA, 16802, USA}
\affiliation{Center for Exoplanets and Habitable Worlds, 525 Davey Laboratory, The Pennsylvania State University, University Park, PA, 16802, USA}

\author[0000-0001-5000-1018]{Andrew J. Metcalf}
\affiliation{Space Vehicles Directorate, Air Force Research Laboratory, 3550 Aberdeen Ave. SE, Kirtland AFB, NM 87117, USA}
\affiliation{Time and Frequency Division, National Institute of Standards and Technology, 325 Broadway, Boulder, CO 80305, USA}
\affiliation{Department of Physics, University of Colorado, 2000 Colorado Avenue, Boulder, CO 80309, USA}

\author[0000-0002-0048-2586]{Andrew Monson}
\affiliation{Department of Astronomy \& Astrophysics, 525 Davey Laboratory, The Pennsylvania State University, University Park, PA, 16802, USA}
\affiliation{Center for Exoplanets and Habitable Worlds, 525 Davey Laboratory, The Pennsylvania State University, University Park, PA, 16802, USA}

\author[0000-0003-0149-9678]{Paul Robertson}
\affiliation{Department of Physics and Astronomy, The University of California, Irvine, Irvine, CA 92697, USA}

\author[0000-0002-4046-987X]{Christian Schwab}
\affiliation{Department of Physics and Astronomy, Macquarie University, Balaclava Road, North Ryde, NSW 2109, Australia}

\author[0000-0002-4788-8858]{Ryan C Terrien}
\affiliation{Carleton College, One North College St., Northfield, MN 55057, USA}

\author[0000-0001-6160-5888]{Jason T.\ Wright}
\affiliation{Department of Astronomy \& Astrophysics, 525 Davey Laboratory, The Pennsylvania State University, University Park, PA, 16802, USA}
\affiliation{Center for Exoplanets and Habitable Worlds, 525 Davey Laboratory, The Pennsylvania State University, University Park, PA, 16802, USA}
\affiliation{Penn State Extraterrestrial Intelligence Center, 525 Davey Laboratory, The Pennsylvania State University, University Park, PA, 16802, USA}

\correspondingauthor{Shubham Kanodia}
\email{shbhuk@gmail.com}

\begin{abstract}
The Habitable zone Planet Finder (HPF) is a fiber fed precise radial velocity spectrograph at the 10 m Hobby Eberly Telescope (HET). Due to its fixed altitude design, the HET pupil changes appreciably across a track, leading to significant changes of the fiber far-field illumination. HPF's fiber scrambler  is designed to suppress the impact of these illumination changes on the radial velocities-- but the residual impact on the radial velocity measurements has yet to be probed on sky. We use GJ 411, a bright early type (M2) M dwarf to probe the effects of far-field input trends due to these pupil variations on HPF radial velocities (RVs). These large changes ($\sim$ 2x) in pupil area and centroid present a harsh test of HPF's far-field scrambling. Our results show that the RVs are effectively decoupled from these extreme far-field input changes due to pupil centroid offsets, attesting to the effectiveness of the scrambler design. 
This experiment allows us to test the impact of these changes with large pupil variation on-sky, something we would not easily be able to do at a conventional optical telescope. While the pupil and illumination changes expected at these other telescopes are small, scaling from our results enables us to estimate and bound these effects, and show that they are controllable even for the new and next generation of RV instruments in their quest to beat down instrumental noise sources towards the goal of a few \cms{}.
\end{abstract}


\section{Introduction} \label{sec:intro}

Precise radial velocity (RV) measurements have demonstrated their utility in both discovery of exoplanets as well as in measuring the masses of transiting planets. Improving this RV technique is critical to enable discovery of terrestrial-mass planets in or near the Habitable zones of their host stars \citep{kopparapu_revised_2013}. Knowing which star to preferentially look at may prove critical for the design and execution of proposed transmission spectroscopy and direct imaging missions such as \added{JWST \citep{greene_characterizing_2016},} LUVOIR \citep{the_luvoir_team_luvoir_2019} and HABEX \citep{gaudi_habitable_2019}, which  greatly benefit from precise planetary mass measurements \citep{batalha_precision_2019} and orbital parameters for both transiting and directly imaged planets. The first exoplanet discovered around a solar type star---51 Peg b---was discovered using precise RVs \citep{mayor_jupiter-mass_1995} obtained using the ELODIE spectrograph \citep{baranne_elodie:_1996}.

\begin{figure*}[!t] 
\centering
\includegraphics[width=\textwidth]{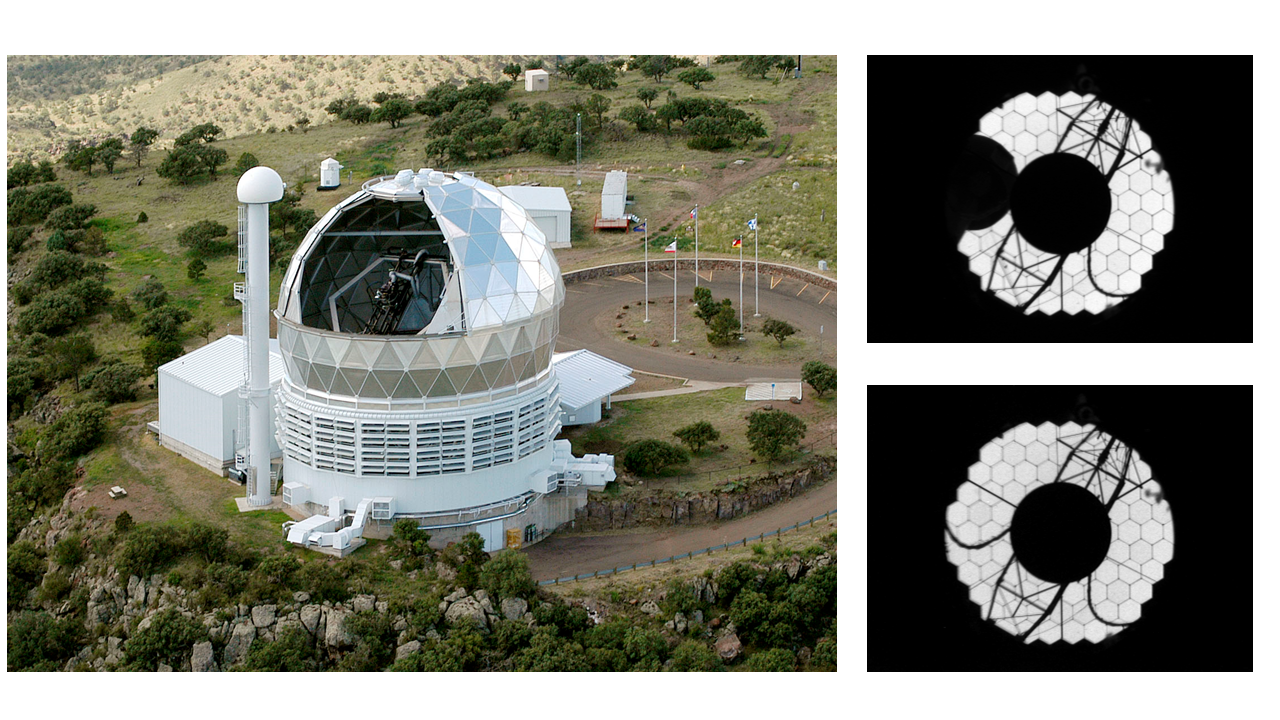}
\caption{\textbf{Left: }An image of HET showing the dome housing the telescope, as well as the CCAS tower to the left of the dome. Credit: Marty Harris/McDonald Observatory. \textbf{Right:} Snapshots of the pupil viewing camera under afternoon sky illumination. The central obscuration with the PFIP, as well as the support strusses. Also visible are the fiber bundles on both sides of the central obscuration. The CCAS tower is visible in the top right image obscuring part of the pupil.} \label{fig:het}
\end{figure*}

ELODIE was one of the  the first fiber fed spectrographs used for precise radial velocity observations. \cite{brown_high_1990} used the Penn State Fiber Optic Echelle \citep[FOE;][]{ramsey_penn_1985} to publish some of the first precision fiber coupled RVs studying p-mode oscillations on Procyon.  \deleted{Switching from slit to fiber fed illumination has multiple advantages. First, it allows the instrument to be decoupled from the telescope structure. This enables a static gravitational vector on the instrument optics, one that does not change with the telescope's position. It also allows for bigger and bulkier instruments with extensive environmental stabilization \citep{lovis_exoplanet_2006, stefansson_versatile_2016}. Secondly, replacing direct slit illumination with optical fibers enables spatial scrambling which can alleviate telescope pointing errors and jitter.}\replaced{There}{Since then, there} has been extensive development of optical fibers for the purposes of increasing their illumination stability, and mitigating the various sources of illumination noise that they add. Optical fibers help desensitize the output flux distribution from changes in the telescope and focal illumination, and offer extensive azimuthal scrambling, but only incomplete radial scrambling \citep{angel_very_1977, avila_photometrical_2006, avila_optical_2008}. Therefore, as the input illumination changes due to guiding errors, seeing, and telescope pupil variations, the output intensity distribution can vary. For fiber fed precision RV instruments, these variations in fiber output illumination pattern can cause un-calibratable spurious RV noise \citep{pepe_harps_2008, fischer_state_2016}. The use of a double scrambler in the fiber feed to exchange the near and far-field illumination\footnote{For an introduction to the importance of far-field and near-field stability patterns, refer to \cite{hunter_scrambling_1992}.} using a lens relay is one way to mitigate this problem \citep{brown_high_1990, hunter_scrambling_1992}. The use of non-circular fiber cores has also been explored, to good effect, to improve the spatial scrambling \citep{chazelas_new_2010, avila_frd_2012, spronck_extreme_2012}.

\begin{figure*}[!t]
\gridline{\href{https://youtu.be/WlmRKfB0D5U}{\fig{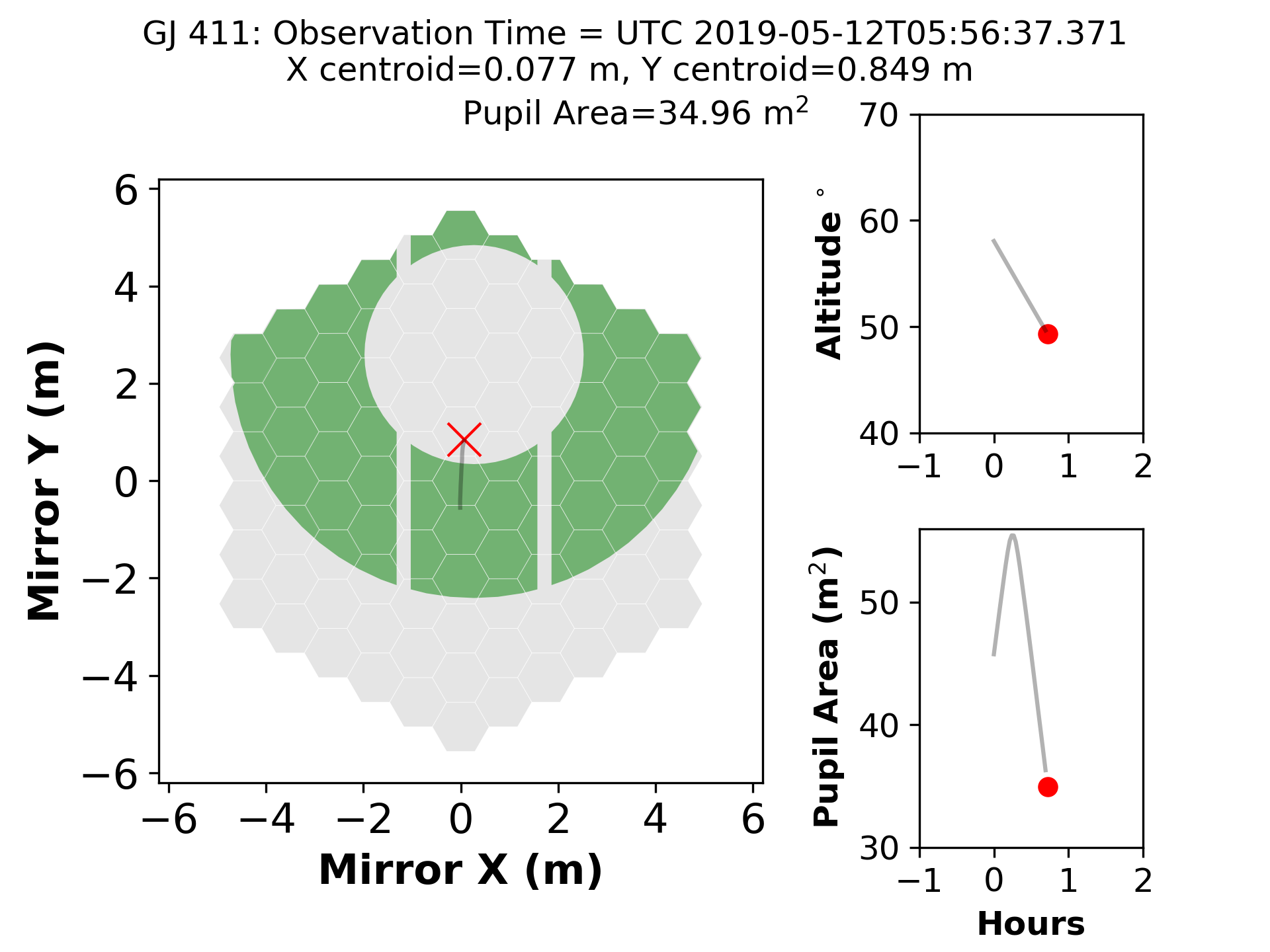}{0.45\textwidth}{\small a) West track observations - 2019 May 12}} 
          \href{https://youtu.be/17K7lDKT8Pk}{\fig{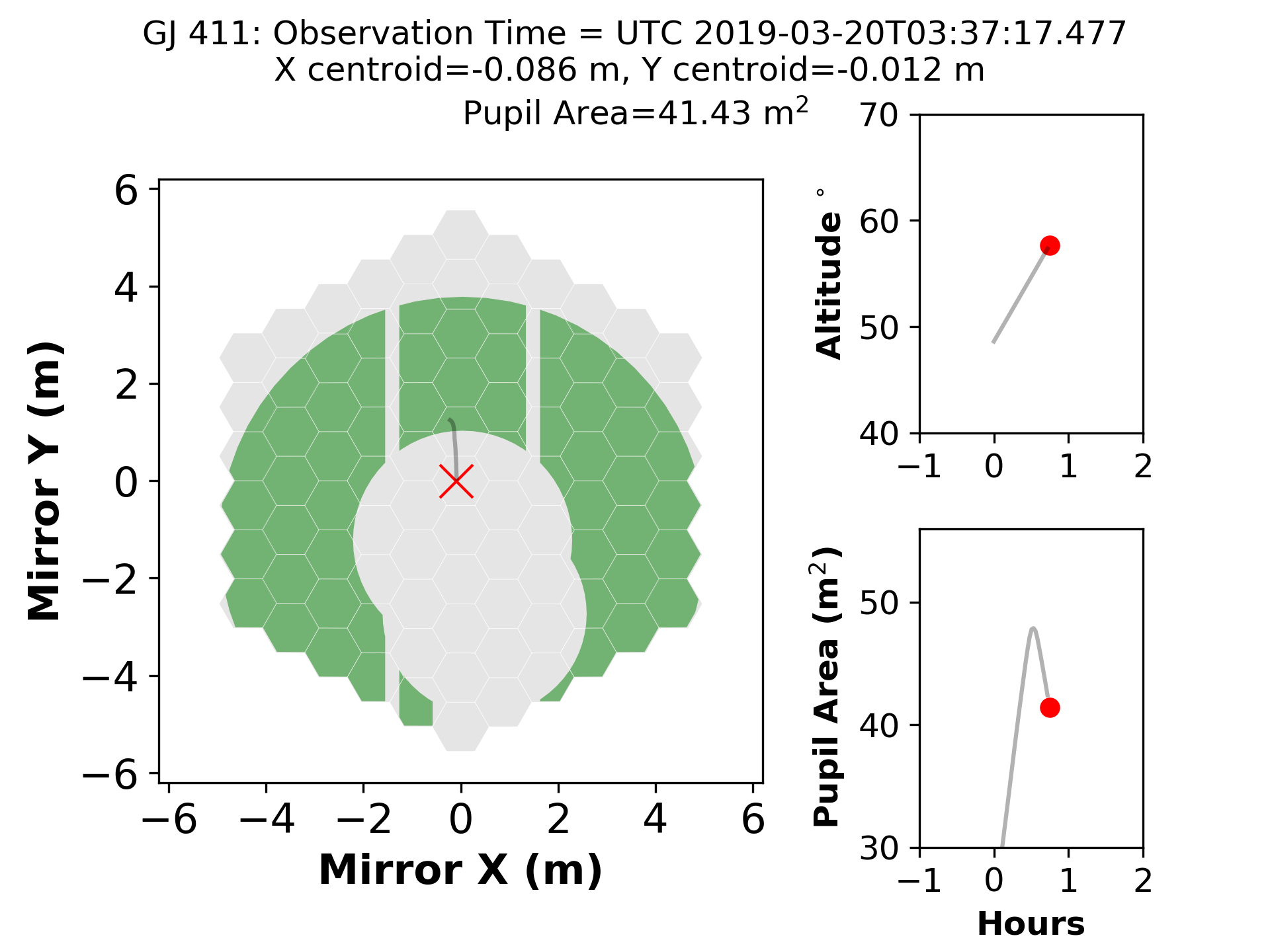}{0.45\textwidth}{\small b)  East track observations - 2019 March 20}}} 
\caption{\small Simulations showing the evolution of the HET pupil for GJ 411 as a function of time for the West and East track separately. In both cases, the left panel shows the circular HET pupil on the hexagonal primary mirror. The grey background shows the telescope primary, where green shows the varying effective pupil illuminated by star light at different epochs. The red cross marks the centroid of the pupil, while the grey line marks its evolution with time. The circular obscuration in the middle is the shadow cast by the Wide-Field corrector (WFC), whereas the two vertical bars mark the support structure. In the case of the East track, we also see the shadow of the CCAS tower (circular + cylindrical obscuration). The CCAS tower is at an azimuth of $\sim 69^{\circ}$ and is used to align the individual primary mirror segments. The right top panel is the altitude of the object as it changes within the track, whereas the bottom right panel is the pupil area changing with time. An animation of these track changes is available on YouTube by clicking on the images. It can be seen that the effective HET pupil (and hence the illuminated portion of the primary) changes in both shape and area as a function of time due to its fixed altitude design. The snapshot shows one of the extremes of the track. \textbf{a)} The West track animation is created from a visit spanning about 45 minutes ($\sim$ 30 minutes exposure + overhead) on 2019 May 12, where each frame represents an individual exposure of 63 seconds exposure. \textbf{b)} Similarly, the East track animation is created from a visit from 2019 March 20 spanning a similar duration. The figure can also be clicked-on to be redirected to the animation.  
}\label{fig:HETpupil}
\end{figure*}

 Motivated by the illumination stability requirements for the near-infrared Habitable-zone Planet Finder (HPF) instrument (Section \ref{sec:hpf}) on the Hobby-Eberly telescope (HET), as well as the availability of refractive index $\sim 2.0$ glasses, \cite{halverson_efficient_2015} adapted the double scrambler into a ball lens double scrambler, which offers excellent scrambling gain\footnote{Scrambling gain is a measure of the output flux variation relative to the variation in input illumination.} ($> 10,000$) and high throughput \added{(85-87$\%$\footnote{From laboratory testing \citep{halverson_efficient_2015}.})} in a compact arrangement. The HET is a  fixed-altitude telescope, where the telescope truss remains fixed in place during an astronomical observation. The prime focus instrument package (PFIP), located about 36 m from the spherical primary, tracks the target in 6 axes across the focal plane of the spherical primary.  Therefore, the effective telescope pupil is defined by the HET primary, as well as the PFIP, and changes in both size and shape continuously through the observation window \citep[\autoref{fig:HETpupil};][]{lee_analysis_2010, lee_facility_2012}.  Without sufficient scrambling, this pupil variation, which manifests as a change in the far-field illumination to the fiber input, would introduce large uncalibratable RV errors unsuitable for precision RV studies. In this paper we present a harsh test for the mitigation of this extreme pupil variation using the ball lens double scrambler \citep{halverson_efficient_2015} using on-sky RVs for the early type (M2) M dwarf GJ 411. Using these RVs we demonstrate the high level of scrambling achieved even in this extreme scenario- and constrain upper limits on the dependence of the RVs on the input pupil of HET. 
 
%  This also validates the scrambling design for NEID, an extreme precision optical RV spectrograph \citep{schwab_design_2016}, which adopts a fiber feed similar to that of HPF.

As the new generation of precision RV instruments transitions to sub \ms{} instrument precision levels in search of the elusive Earth analogues inducing a Doppler RV semi-amplitude of lesser than 10 \cms{}, it will be important to ensure excellent input illumination stability and fiber scrambling, even with more conventional telescope designs. Changes in the incident near and far-field illumination patters due to guiding errors, residual atmospheric chromatic dispersion, and atmospheric effects, directly lead to systematic variations in the spectrometer point-spread-function (PSF). Furthermore, for instruments currently being built for large telescopes with segmented mirrors, there could be variations in the pupil from night to night due to reflectivity differences between segments, as well as holes in the pupil when individual segments are removed. For the next generation instruments aiming for 30 \cms{} or better precision \citep{jurgenson_expres_2016, schwab_design_2016,pepe_espresso:_2014}, illumination variations typically need to be stable at the few \cms{} level, considering the family of error sources affecting the measurements \citep{halverson_comprehensive_2016}.  In Section \ref{sec:het} we discuss the structure of HET, in Section \ref{sec:hpf} we detail HPF's fiber train as well as the ball lens double scrambler. The HPF RVs used in this analysis are explained in \ref{sec:rvs}, whereas in Section \ref{sec:expdesign} we list the assumptions made, and justification for the analysis which is detailed in Section \ref{sec:analysis}. Finally we summarize our results our results in Section \ref{sec:conclusion}, and discuss the relevance to other precision RV instruments.

\section{The Hobby-Eberly Telescope}\label{sec:het}
The HET has a 11 m hexagonal shaped {\it spherical} primary mirror, with a pupil diameter of 10 m, and consists of 91 hexagonal segments of 1 m each \citep{ramsey_early_1998,   hill_current_2012}. It has a central obscuration 4.5 m in diameter \citep{booth_wide_2006}, is fixed at an altitude of $55^{\circ}$ and can rotate in azimuth; whereas the PFIP can track objects from altitudes ranging from 48$^{\circ}$ to 65$^{\circ}$. Located in the PFIP is the focal plane array (FPA) \citep{lee_facility_2012}, which consists of the telescope fibers for HPF \citep{kanodia_overview_2018},  and the multiple integral field units (IFUs) for the Low Resolution Spectrograph \citep[LRS2;][]{lee_lrs2:_2010}, and the HET Dark Energy Experiment \citep[HETDEX; ][]{hill_virus:_2012} VIRUS instrument. HET is a fully queue-scheduled telescope with all observations executed in a queue by the HET resident astronomers \citep{shetrone_ten_2007}. To align the individual primary segments on this fixed altitude telescope, a center of curvature alignment sensor (CCAS) is located at the center of curvature of the primary mirror atop a 28 m tower situated approximately 15 m away from the telescope at azimuth 68.59$^{\circ}$ (\autoref{fig:het}). Given the pointing restrictions of the HET a bright star may not always be available at the beginning of the night for mirror segment alignment, and the CCAS tower enables deterministic alignment every night.

\begin{figure*}
\gridline{\fig{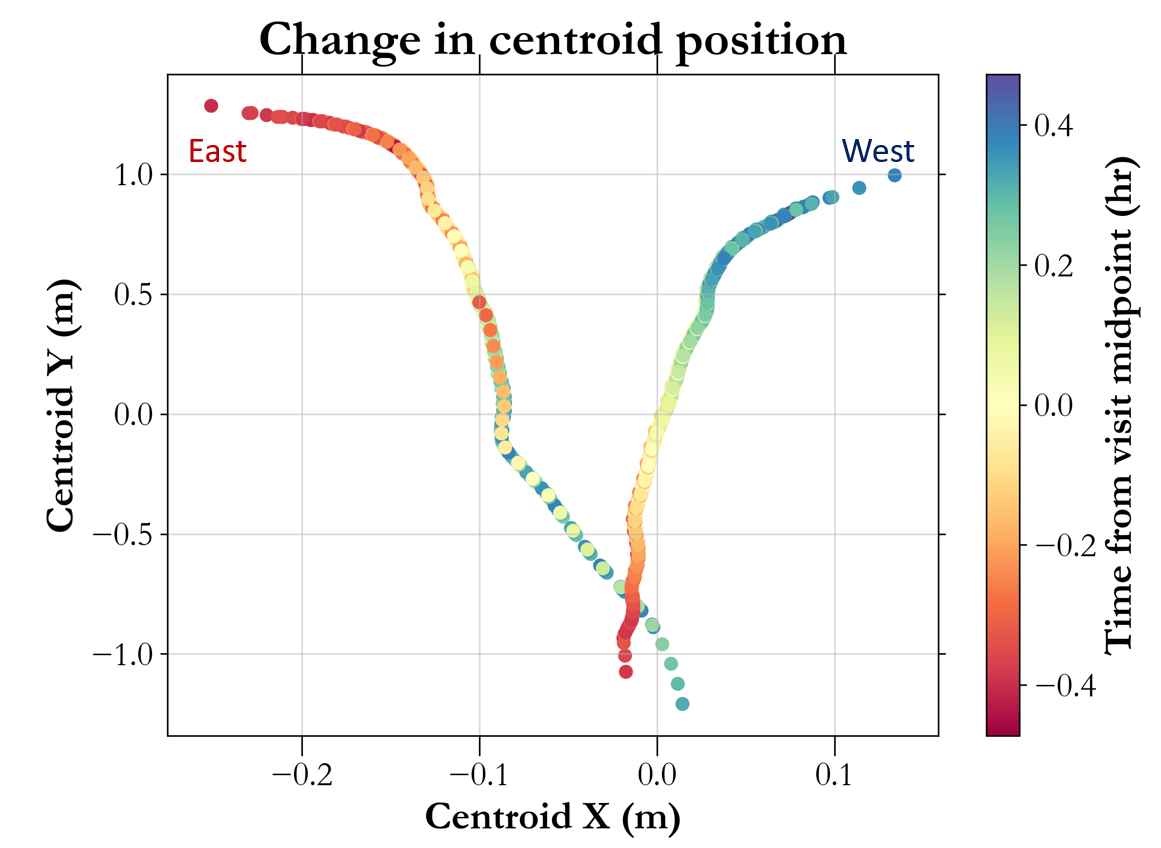}{0.45\textwidth}{{\small a)  Change in centroid across track.}\label{fig:Centroid}}    
          \fig{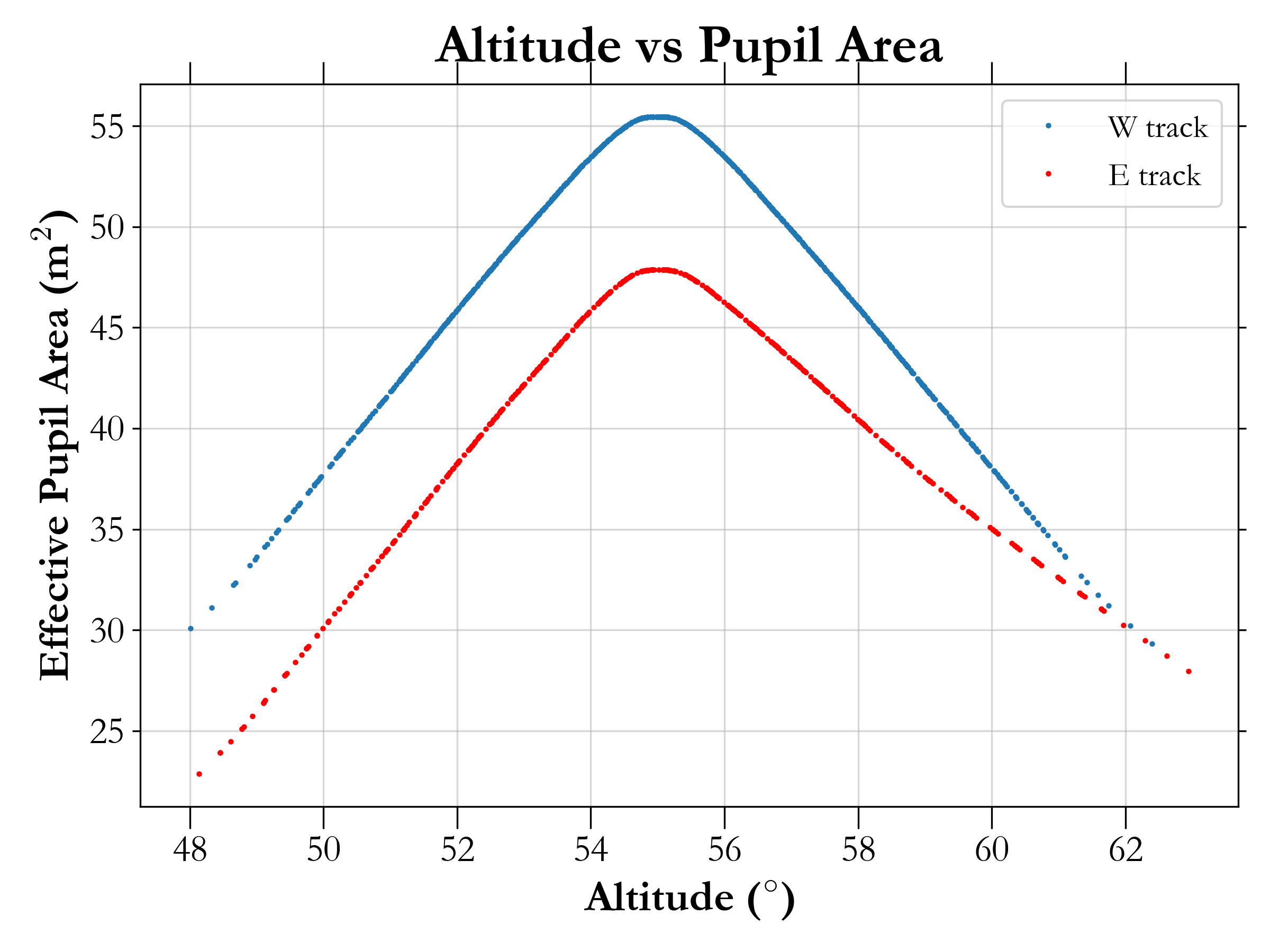}{0.42\textwidth}{ \small b)  Change in Altitude vs Pupil Area}\label{fig:alt_vs_area}} 
\caption{\small \textbf{a)} The pupil centroid position for each exposure as a function of time within each track. These centroids have been estimated using a simulated pupil image (\autoref{fig:HETpupil}), which includes the shadow cast by the WFC and the CCAS tower. The colour represents when in each track the exposure was taken. For example yellow represents 0 or the midpoint of the visit. Each observation window typically lasts about an hour, during which the visit is of $\sim$ 30 minutes duration, and consists of 30 exposures. The two different lines represent the East and West observing tracks for HET, and are asymmetric because the East track observations include a shadow from the CCAS tower, which is not present in the West track. Note the change in scales between x and y axes. The normalized offset is the pupil centroid offset divided by the pupil diameter of 10 m. \textbf{b)} The effective pupil area for HET as a function of altitude for all 1211 exposures. As expected, the telescope pupil is maximum when the mirror and WFC are perfectly aligned, and the object altitude is 55$^{\circ}$. The blue curve represents the West track, while the red one is the East track. The West track attains a higher maximum area since the pupil for these observations is not occulted by the CCAS tower, whereas the range of azimuths for the East track do include the CCAS tower shadow. The centroid and pupil area shown in this figure were calculated using the simulation shown in \autoref{fig:HETpupil}.}\label{fig:PupilChange}
\end{figure*}

The combination of altitude and azimuth tracking, typically presents observation windows of $\sim 1$ hour\footnote{The exact duration depends on the declination of the target.} duration per visit, commonly referred to as a track. Depending on the azimuth and declination of the target, there are East and West tracks of observability (shown in \autoref{fig:PupilChange}a). Within this track, each HPF visit analyzed in this work typically lasts 45 minutes for GJ 411 ($\sim$ 30 minutes exposure + overhead), and consists of multiple exposures (Section \ref{sec:rvs}). If the object is continuously observed across the entire duration of the track, the pupil area can change by as much as a factor of 2 between the track extrema and its median position\footnote{This is equivalent to a change in the effective pupil diameter from $\sim 8.5$ m to 5.5 m.}. The pupil is obscured by the WFC in the center, two support structure on each side of the obscuration, as well as  fiber bundles with the instrument fibers. In addition, the CCAS tower obscures part of the pupil for observations in the East track between azimuths 41$^{\circ}$ - 97$^{\circ}$.  We simulate the HET pupil using the \texttt{python} package \texttt{pyHETobs} which is further discussed in  \autoref{sec:pyHETobs}. In \autoref{fig:HETpupil}, we show a simulation of the pupil change across the track for an observation of GJ 411. \autoref{fig:PupilChange} shows the change in centroid for each exposure within the tracks, as a function of when the exposure was taken within the track. This drastic change in the effective pupil area, centroid position, and shape presents a change in the far-field input to the fiber, which necessitates excellent spatial scrambling for precision radial velocities.

\section{HPF and the HPF Ball scrambler}\label{sec:hpf}
\subsection{Overview of HPF}
The Habitable zone Planet Finder \citep[HPF;][]{mahadevan_habitable-zone_2012, mahadevan_habitable-zone_2014}, at the 10 m Hobby Eberly Telescope \citep[HET;][]{ramsey_early_1998, booth_hobby-eberly_2003, booth_hobby-eberly_2004} is a near-infrared (NIR) fiber fed precision RV spectrograph capable of simultaneous observations with a science, sky and simultaneous calibration fiber \citep{kanodia_overview_2018}. With active environmental stability control, it achieves $\sim 1 ~ \rm{mK}$ temperature stability \citep{stefansson_versatile_2016, robertson_system_2016}.  HPF covers the  0.808 to 1.280 $\mu$m wavelength region, was deployed at HET in October 2017, and started full science operations in mid 2018. With an on-sky demonstrated RV performance of 1.5 \ms{} on Barnard's star over an extended baseline, it is currently the most precise RV instrument in the NIR  \citep{metcalf_stellar_2019}. HPF uses the ball lens double scrambler \citep{halverson_efficient_2015} arrangement in order to achieve the highly stable input illumination required for precise RV measurements. For HPF we placed a formal requirement of on the illumination stability, equivalent to 30 cm s$^{-1}$ of associated RV error, and includes both near-field (guiding errors, and pointing jitter) as well as far-field changes (pupil changes). A full overview of the HPF fiber system is described in \cite{roy_scrambling_2014} and \cite{kanodia_overview_2018}, though we summarize the top-level design details here.

\begin{figure*}[!t] 
\centering
\includegraphics[width=0.8\textwidth]{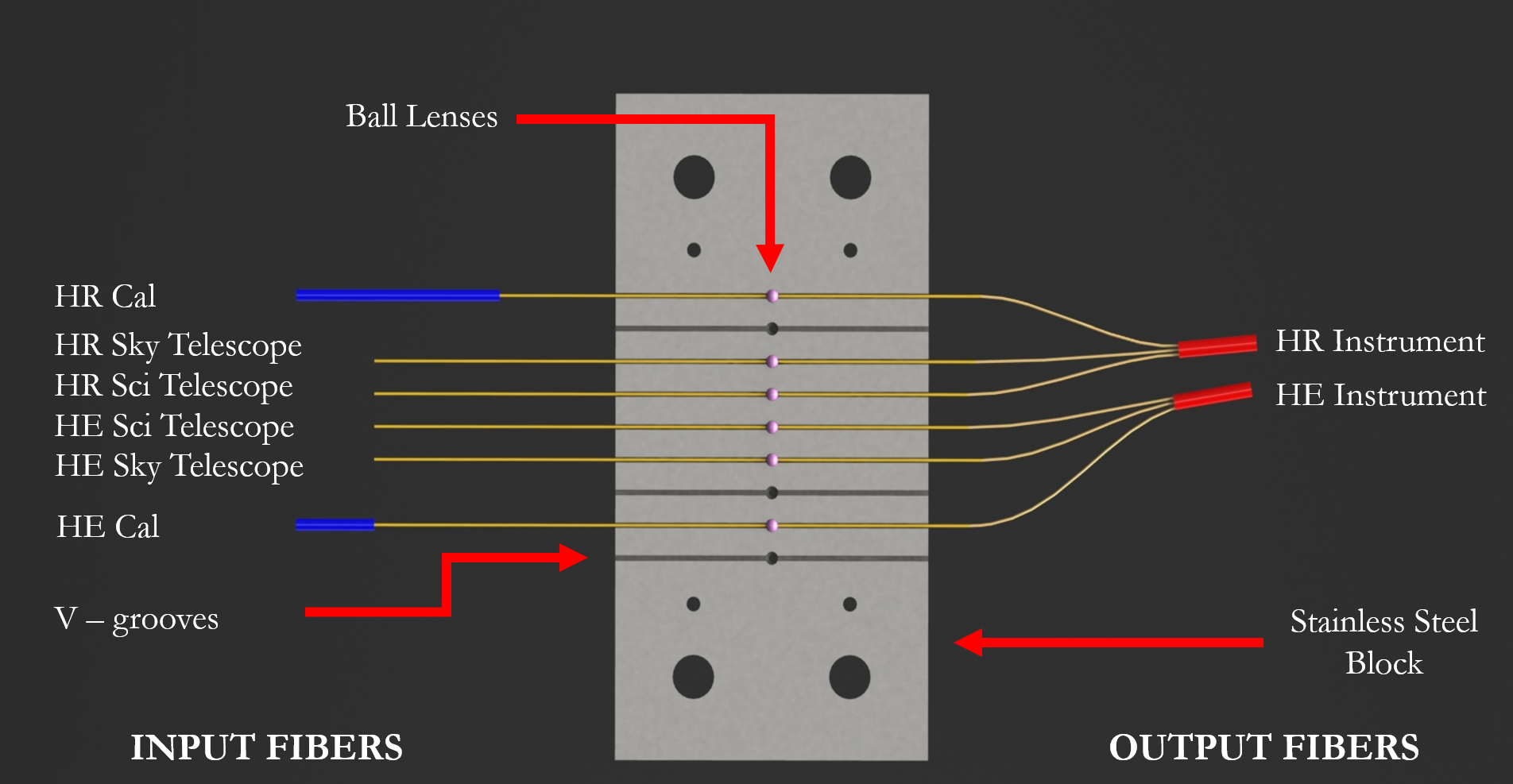}
\caption{Schematic of the HPF v-groove block with the ball lenses and octagonal fibers. The input (telescope + calibration) fibers for the High Resolution (HR) and High Efficiency (HE) mode come in from the left. The output fibers are on the right, where the two bundles are for HR and HE. The output fiber bundles are then spliced on to circular fibers which serve as the instrument input. We use the HR mode for all GJ 411 observations.} \label{fig:vgroove}
\end{figure*}

\begin{figure}[!t] 
\centering
\includegraphics[width=\columnwidth]{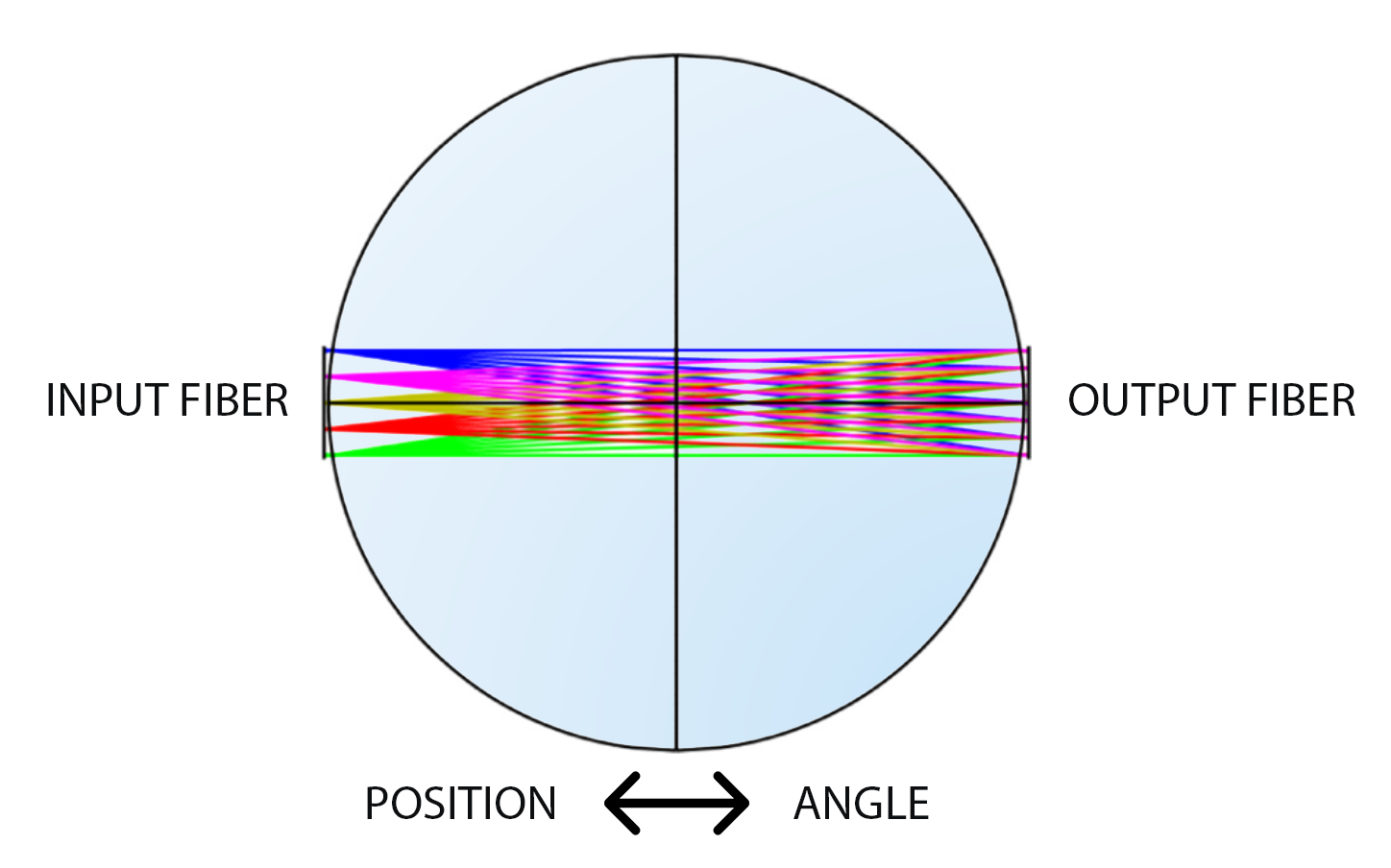}
\caption{Adapted from \cite{halverson_efficient_2015}, we show the interchange of near and far field illumination, i.e., conversion of position to angle.} \label{fig:balllens}
\end{figure}

\subsection{HPF Fiber delivery system}\label{sec:hpffiber}
HPF uses a combination of octagonal and circular fibers, and a single-element spherical optical double scrambler. The fibers provide high levels of near-field (image plane) scrambling, while the double scrambler is used to homogenize the far-field (pupil plane) illumination by exchanging the near and far field of the two fibers coupled to the double scrambler. The near-field of the output fiber is imaged on to the detector, and represents the positional intensity distribution on the fiber face at a given wavelength. The far-field is the angular distribution of the fiber output and is in the pupil plane, as projected on to the grating. Changes in the \added{spectrometer} pupil plane due to imperfect far-field scrambling manifest as varying illumination of the grating. These changes can cause spurious RV shifts when coupled with grating inhomogeneities as well as wavefront error. Imperfect far-field scrambling would also cause changes in the illumination of the spectrograph optics, which could manifest as shape changes of the PSF.

The HPF fiber feed uses a high refractive index (n $\sim 2$)\footnote{Therefore the focal length of the ball is equal to it's radius, ensuring that the image is formed at (or very close to) the surface of the lens (\autoref{fig:balllens})} ball lens to efficiently image the far-field illumination of the input fiber (telescope) onto the output fiber, producing a smooth, scrambled pattern in both near and far-field. This near-field of the output fiber bundle is further azimuthally scrambled by the fiber and the output (spectrograph end) is then imaged on to the HPF detector plane. The input fiber (at the prime focus of the telescope) is an octagonal core fiber of core diameter 299 $\mu$m, which is fed to a stainless steel (420 SS) block with v-grooves, where the fiber is face-coupled to an anti-reflection coated, 2.0 mm S-LAH79 ball lens (\autoref{fig:vgroove}). The grooves in the stainless steel block are precisely machined using electrical discharge machining.  The output end consists of a 2 meter patch of octagonal fiber which is then spliced on to a circular core fiber of diameter 312 $\mu$m \citep{kanodia_overview_2018} to further improve the scrambling performance.  The 2 meter fiber section is mechanically agitated to mitigate modal noise \citep{halverson_modal-noise_2015, kanodia_overview_2018}.  %The change in input pupil to the telescope fiber therefore is converted to a near-field change for the output fiber. If uncorrected, this would manifest as a change in the image formed of the fiber on the detector (focal plane). 

\subsection{Lab Tests of double scrambler}\label{sec:lab}

\begin{figure}[!t] 
\centering
\includegraphics[width=\columnwidth]{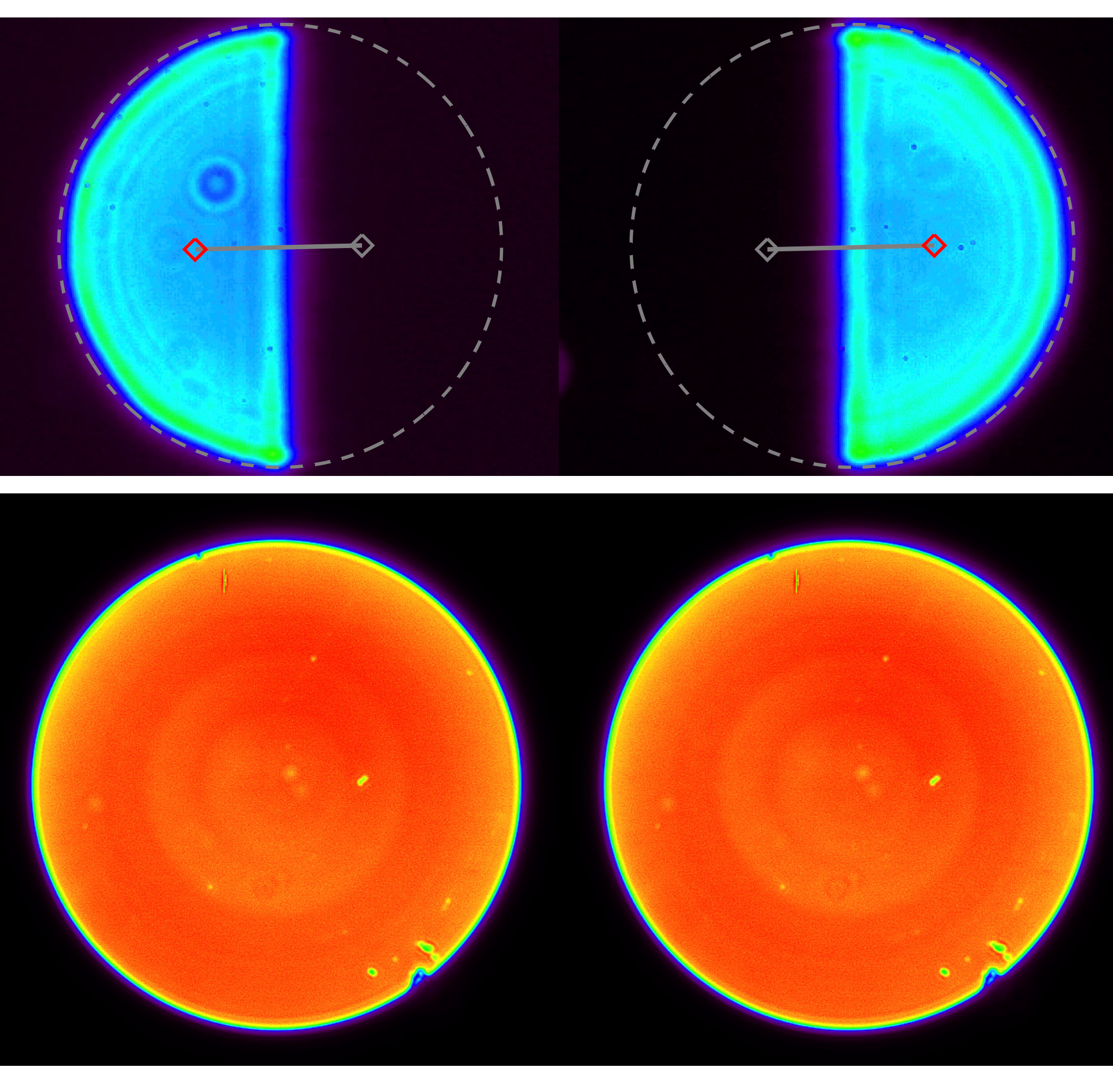}
\caption{Using the lab setup described in \citep{halverson_efficient_2015}, we show the dependence of the output near-field on the input pupil variations. The fiber setup is similar to \autoref{fig:vgroove} where the input octagonal fiber feeds a ball lens, followed by another octagonal fiber which is spliced on to the output circular fiber. \textbf{Top:}  The input pupil with the pupil mask. \textbf{Bottom:} shows the near-field of the circular fiber. The input pupil represents a 40$\%$ change in pupil centroid position with respect to the pupil diameter, whereas the change in centroid for the output near-field corresponds to an RV change of $61 \pm 2.5$ \cms{}. This is equivalent to a slope of 1.54 $\pm$ 0.06 \ms{} per unit change in normalized pupil centroid offset.} \label{fig:halfmoonpupil}
\end{figure}

To gauge the sensitivity of the HPF fiber system to input illumination variations, we measured the output illumination pattern using the laboratory measurement apparatus described in \cite{halverson_efficient_2015}.  A broadband, fiber-coupled source was used to inject light into a prototype fiber system that emulated the configuration of the final HPF fiber train. To specifically probe the sensitivity of the system to incident pupil variations which are expected to dominate the noise floor for HPF due to the intrinsically variable HET pupil, a mask was placed at the pupil prior to the test fiber. To estimate the RV sensitivity of the fiber system output to changes in the incident pupil, we convert the measured variations in the fiber near-field to the effective velocity shift that would be measured within HPF (\autoref{fig:halfmoonpupil}). The pupil centroid between the two inputs shown in \autoref{fig:halfmoonpupil} changes by 40$\%$ of the pupil diameter, and causes a change in the output near-field. We carefully measure the change in centroid of this output near-field, which would correspond to an RV shift of $61 \pm 2.5$ \cms{} for HPF. This is equivalent to a slope of 1.5 $\pm$ 0.06 \ms{} per unit change in normalized pupil offset, i.e. for a 10$\%$ change in centroid position, we would expect an RV offset of $\simeq$ 15 \cms{}. Note that this corresponds to a worst case scenario where the entire near-field movement of the PSF is along the dispersion axis, realistically the near-field movement would be along an axis between the dispersion and cross-dispersion directions, and hence distributed between the two. \added{In addition, the throughput of the ball lens double scrambler was measured using an 830 nm and 1310 nm laser to be between 85-87$\%$.}

We assume a linear relationship for the impact of input illumination offsets on the output near-field, based on the traditionally followed methodology for parameterizing the scrambling gain \citep{avila_optical_2008, avila_frd_2012, halverson_efficient_2015}.  We use this assumption to place upper limits between pupil parameters and GJ 411 RVs (Section \ref{sec:analysis}).

\section{GJ 411 Radial Velocities}\label{sec:rvs}

\begin{figure*}[!t] 
\centering
\includegraphics[width=\textwidth]{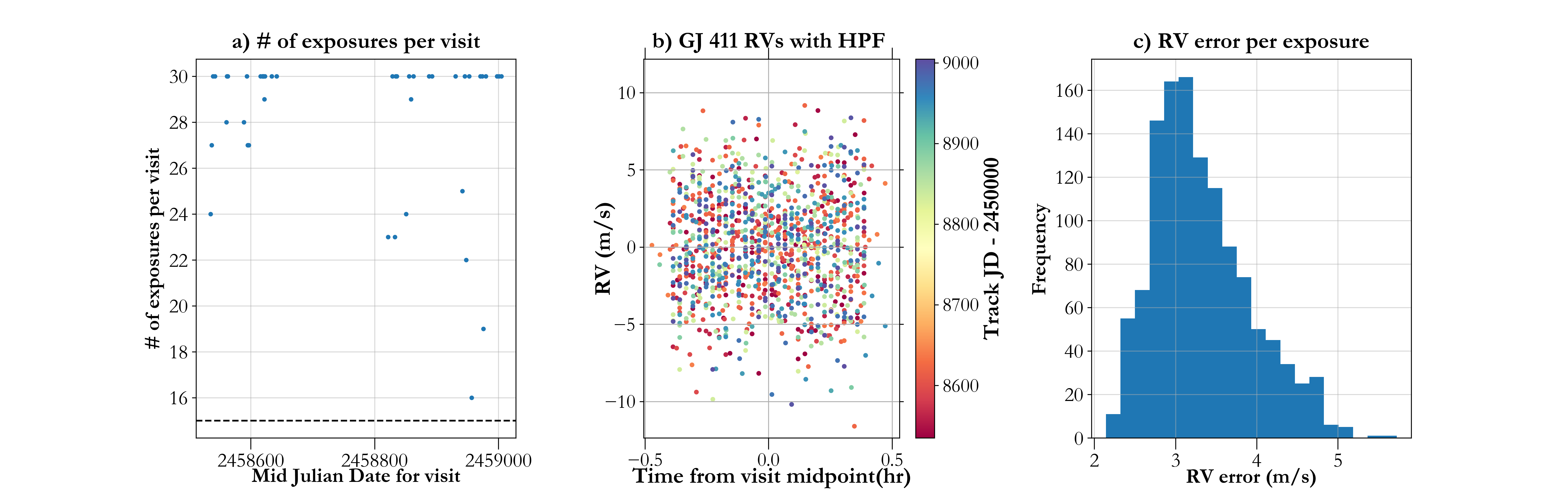}
\caption{\textbf{a)} Showing the number of exposures per HPF visit on GJ 411 as a function of time. We exclude visits where we have fewer than 15 exposures per visit. This excludes a total of 73 exposures across 10 visits. \textbf{b)} HPF RVs for GJ 411 as a function of time, where we subtract the midpoint of the visit from the x axis. The points are colour coded by the Julian Date of the exposures. We do not see any obvious trends in the RVs within each track. Each visit includes up to 30 exposures of 63 seconds each, and an overhead of $\sim 30$ seconds for each exposure. \textbf{c)}  Histogram of the RV errors per exposure (63 seconds) computed using \texttt{SERVAL}, where the mean is 3.3 \ms{}.} \label{fig:RVs}
\end{figure*}

GJ 411 \citep[HD 95735, Lalande 21185;][]{gliese_nearby_1979} is a metal poor \citep[M/H = -0.35;][]{mould_infrared_1978} early M dwarf \citep[M2; ][]{mann_how_2015}. It is the brightest M dwarf \citep[J=4.2; ][]{cutri_vizier_2003} in the northern hemisphere and at a distance of 2.55 parsecs \citep{van_leeuwen_validation_2007}, one of the closest stars to the Sun \citep{henry_solar_1994}. GJ 411 is also reported to have a non-transiting planet in a 12.946 day orbital period \citep{diaz_sophie_2019, stock_carmenes_2020}. This star is routinely observed by HPF as part of its long term RV monitoring and engineering program. Its brightness allows for short HPF exposures of 63 seconds each, and up to 30 individual exposures per visit inside each track\footnote{With an exposure overhead of about 30 seconds between exposures.}. Each of these exposures consist of 6 non destructive readout (NDR) up the ramp (UTR) frames of 10.5 seconds each. As discussed in Section \ref{sec:het}, the effective HET pupil area and shape change across the track, and therefore each of these 30 exposures are taken with a different pupil area and centroid (\autoref{fig:HETpupil}). We use these high cadence RV observations of GJ 411  to conduct a harsh test of the far-field scrambling using the ball lens scrambler on HPF, since they allow us to probe the RV impact of changes in the effective pupil parameters. Since we do not discuss the GJ 411 RVs from an astrophysical perspective, but use them to probe instrumental effects, we do not provide a table of the RVs.

We correct for bias noise, non-linearity correction, cosmic ray correction, slope/flux and variance image calculation using \texttt{HxRGproc} \citep{ninan_habitable-zone_2018}. Following the methodology described in \cite{stefansson_sub-neptune-sized_2020} to derive the RVs, we use a modified version of the \texttt{SpEctrum Radial Velocity AnaLyser} pipeline \citep[\texttt{SERVAL};][]{zechmeister_spectrum_2018}. \texttt{SERVAL} uses the template-matching technique to derive RVs \citep[e.g.,][]{anglada-escude_harps-terra_2012}, where it creates a master template from the target star observations, and determines the Doppler shift for each individual observation by minimizing the \(\chi^2\) statistic. We created an individual template for each HPF visit using all the individual exposures within a track. Comparing the RVs for each track to individual templates, allows us to probe for changes in the RVs within each track separately without any long term stellar or instrumental effects complicating the analysis. To create these templates, we explicitly mask out any telluric regions identified\footnote{We mask everything below 99.5$\%$ transmission as tellurics for this template mask.} using a synthetic telluric-line mask generated from \texttt{telfit} \citep{gullikson_correcting_2014}, a Python wrapper to the Line-by-Line Radiative Transfer Model package \citep{clough_atmospheric_2005}. To perform our barycentric correction, we use \texttt{barycorrpy}, the Python implementation \citep{kanodia_python_2018} of the algorithms from \cite{wright_barycentric_2014} to perform the barycentric correction.

We have 1211 exposures of 63 seconds each, across 43 visits spanning early 2019 to mid 2020. We obtain these after filtering out exposures with signal to noise per resolution element (S/N) lesser than 300\footnote{Calculated at 1100 nm}, and visits with less than 15 exposures within the track (\autoref{fig:RVs} (a))

\section{Experimental Design}\label{sec:expdesign}
As mentioned in Section \ref{sec:rvs}, we use the template matching method using the HPF modified \texttt{SERVAL} pipeline to calculate the RVs for GJ 411, and create individual templates for each track (\autoref{fig:RVs} b). Doing this allows us to correct for any offset introduced due to long term instrumental, telescope and astrophysical RV trends, and probe for correlations in the RVs within the tracks, as a function of various telescope parameters. The potential noise terms due to fiber illumination which can be attributed to the input illumination are divided into near-field, and far-field\footnote{Due to their finite size and efficiency, multimode fibers are the most commonly used type of optical fibers (as opposed to single mode). These fibers suffer from speckling, also called modal noise, due to the finite number of transverse modes that propagate across the fiber \citep{hill_modal_1980, rawson_frequency_1980, goodman_statistics_1981}. The most common mitigation for modal noise involves the temporal agitation of the fibers \citep{baudrand_modal_2001, chen_origin_2006, mccoy_optical_2012, mahadevan_suppression_2014, roy_scrambling_2014, petersburg_modal_2018}.}, and are discussed below:

\begin{enumerate}
  
\begin{figure}[b] 
\centering
\includegraphics[width=\columnwidth]{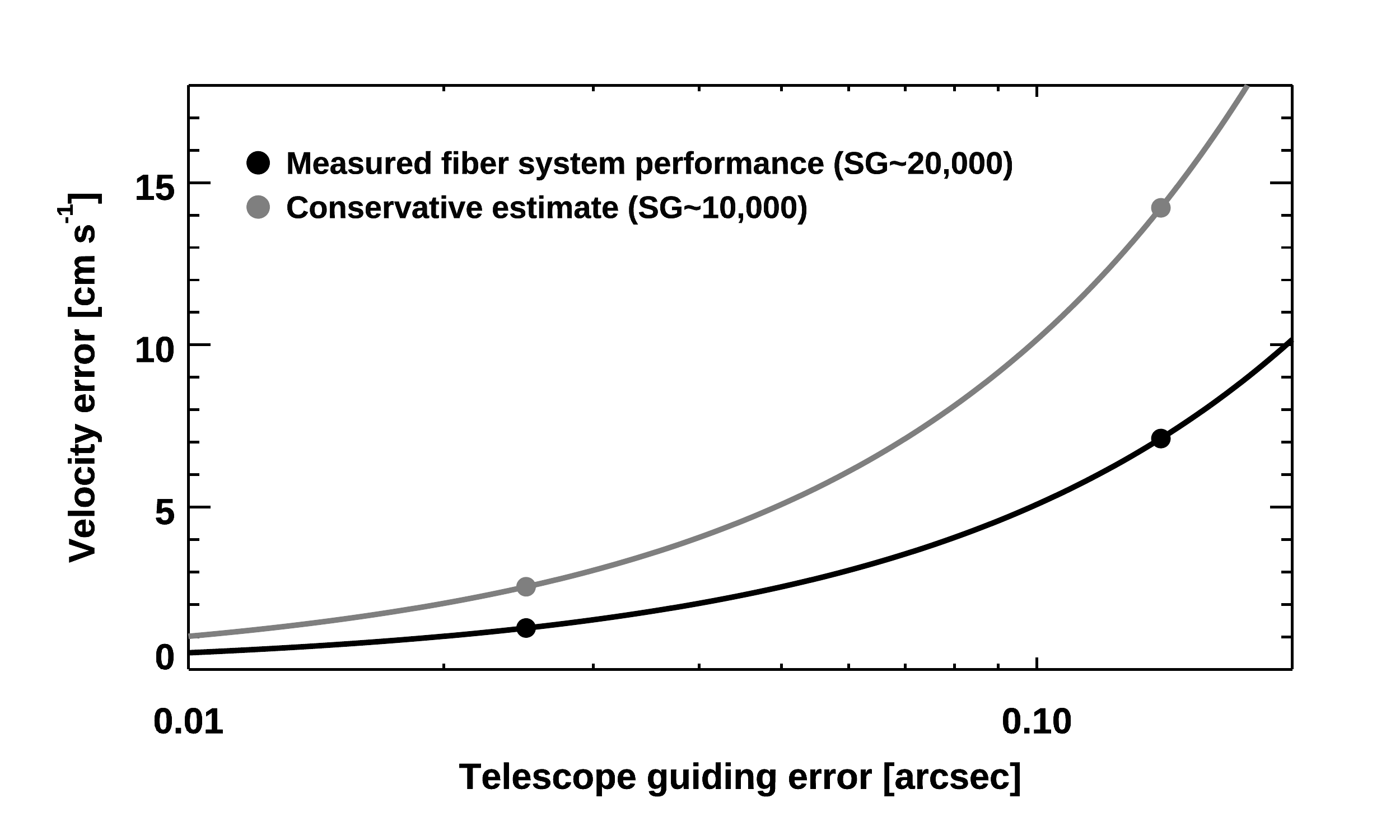}
\caption{Similar to Figure 1 from \citep{halverson_efficient_2015}, the RV error as a function of telescope guiding errors is represented for two different scrambling gains. A conservative estimate of the scrambling gain is 10,000 from lab tests, whereas the upper limit was placed at $\simeq$ 20,000. For individual guide camera exposures at 6.3 second cadence, the RMS guiding is about 0.15 \arcsec which would cause an RV error of $\sim$ 14 \cms{} assuming the conservative scrambling gain. Similarly, when averaged during an HPF exposure of duration 63 seconds, this RMS averages to about 0.025\arcsec, and an RV error of 2 \cms{}.} \label{fig:guidecamjitter}
\end{figure}
     
\begin{figure*}[!t]
\gridline{\fig{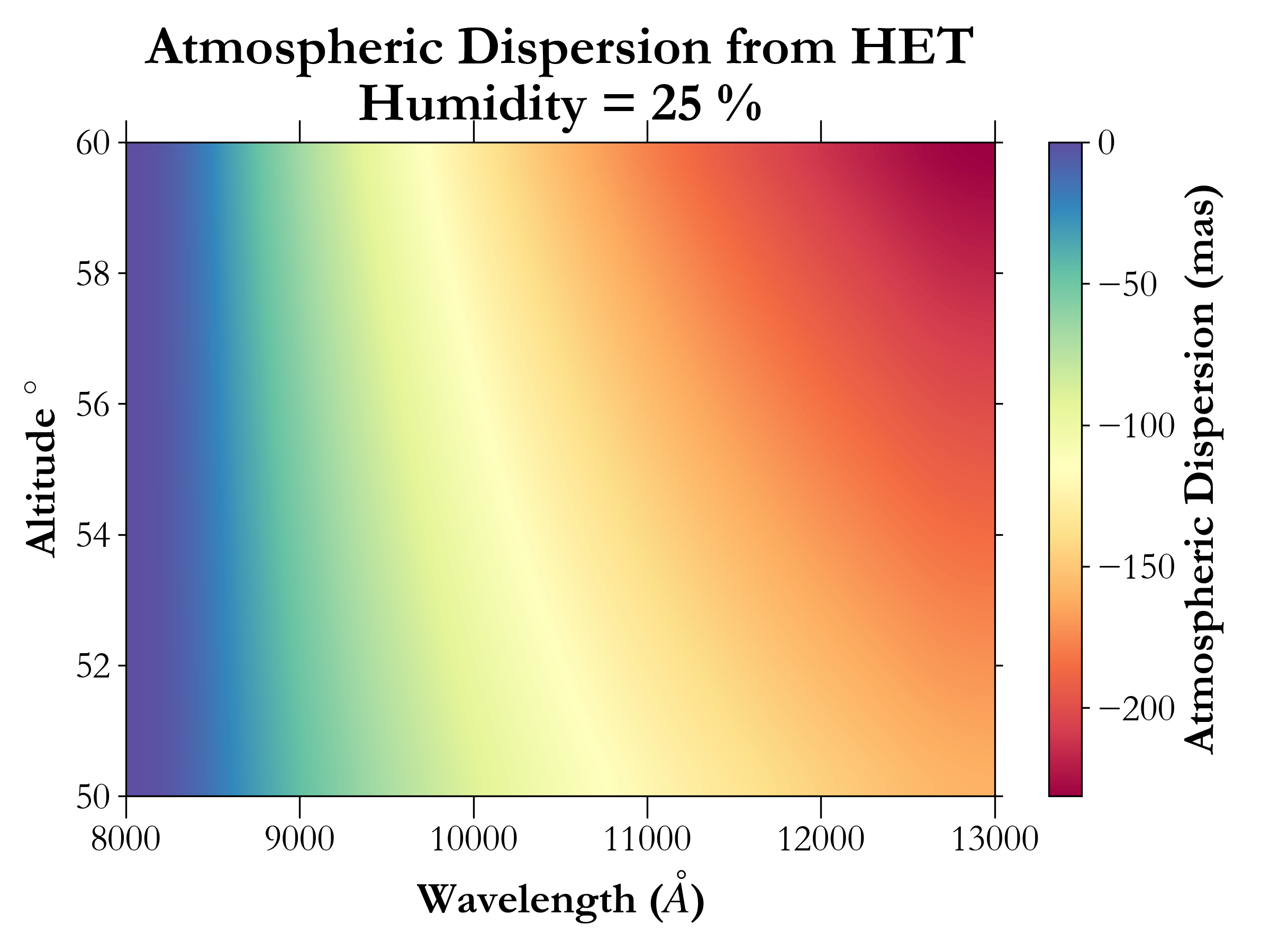}{0.45\textwidth}{{\small a)  Atmospheric Dispersion vs Altitude}}    
          \fig{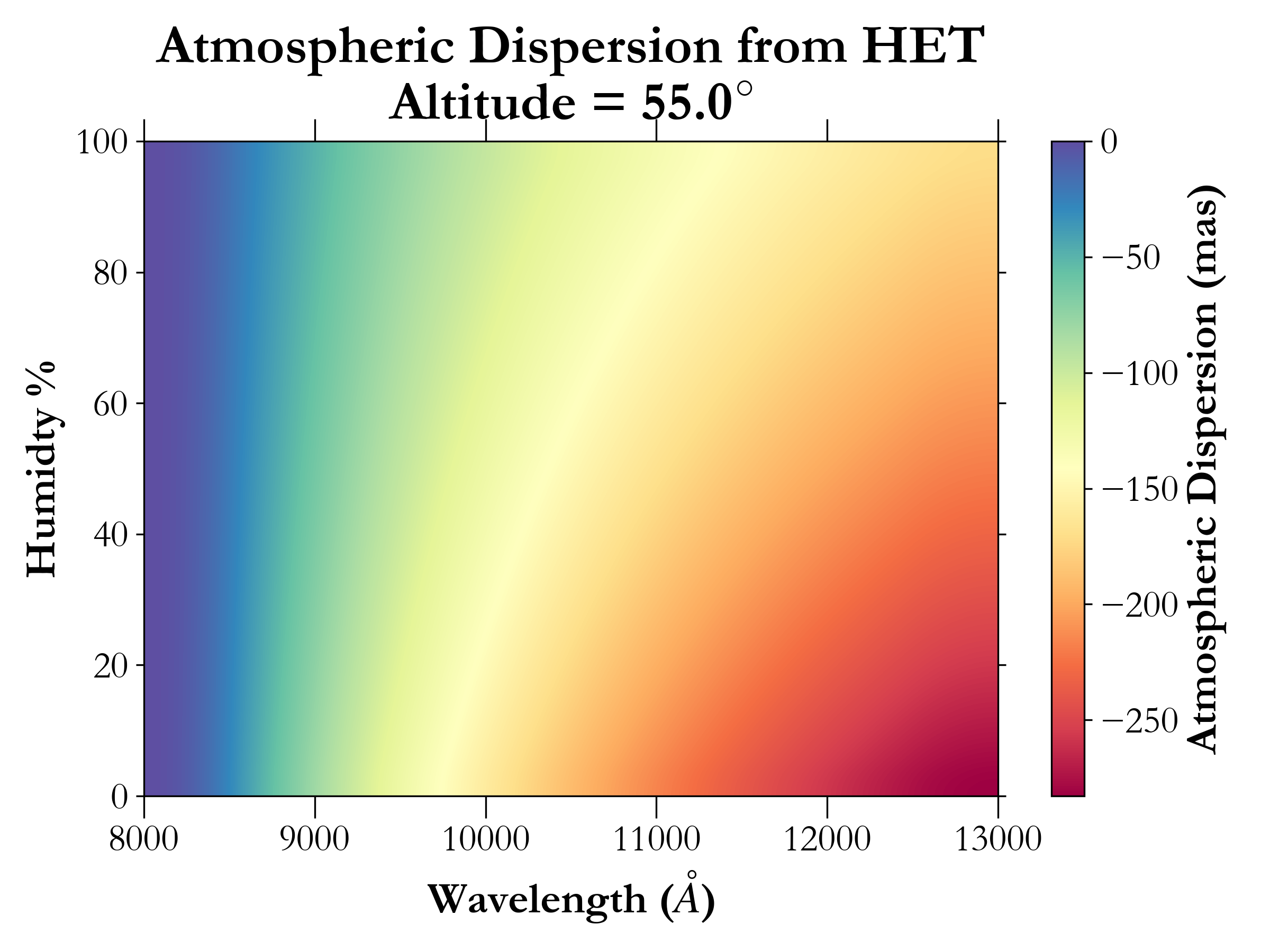}{0.45\textwidth}{ \small b)  Atmospheric Dispersion vs Humidity}} 
\caption{\small The extent of atmospheric dispersion offset from the blue (8200 \AA) to the red (13000 \AA) end of the HPF bandpass as a function of altitude \textbf{(a)}, and humidity \textbf{(b)}. We see that the typical dispersion is $\sim$ 100 mas. The typical humidity for the GJ 411 observations was 17$\%$ to 55$\%$ (16-84$\%$ percentile).}\label{fig:Dispersion}
\end{figure*}

    \item Near-Field illumination: 
    \begin{itemize}
        \item Pointing jitter - Random motion due to wind shake and errors in guiding can add a white noise term to the RVs. We analyze the centroid positions on the HET guide cameras to estimate the guiding RMS to be 0.15\arcsec~unbinned (6.3 second cadence), and about 0.025\arcsec~when binned to the HPF exposure duration for GJ 411 of 63 seconds. Based on the analysis from \cite{halverson_efficient_2015}, we estimate the RV contribution of this to be about 14 \cms{} and 2 \cms{} respectively  (\autoref{fig:guidecamjitter}). However, even if we conservatively assume that the pointing jitter does not bin down, at an RMS jitter of 0.15\arcsec~of jitter, the RV impact is 14 \cms{}; this is substantially lower than our median RV noise of 3.3 \ms{} in 1 minute exposures, and in a quadrature sense is negligible. Hence we do not include the impact of pointing or guiding jitter for this analysis. 
      
        \item Pointing offset - Changes in the near-field input illumination can be caused due to errors in placement of the star on the HPF input fiber, and we estimate this to be of the same magnitude as pointing jitter. Any RV systematics caused due to pointing offsets do not affect this analysis of far-field scrambling, since we subtract a track specific offset from the RVs which accounts for such effects (\autoref{eq:linearfit_mediansubtracted}).

        \item Atmospheric Dispersion - We do not have an atmospheric dispersion corrector (ADC) at HET. Typically this does not cause appreciable error, due to the fixed altitude design of the telescope, and lower amount of atmospheric dispersion in the NIR than the optical. The HPF acquisition camera uses a \replaced{Semrock narrow band filter}{30 nm wide narrowband filter centered around 857 nm} \footnote{\href{https://www.semrock.com/FilterDetails.aspx?id=FF01-857/30-25}{Semrock FF01-857/30-25}} to acquire the star and centroid it on the 1.7\arcsec~HPF fiber in a repeatable manner. Even with changing weather conditions (humidity, pressure, etc.), as well as altitude, the changes in the atmospheric dispersion are on the order of 100 mas (\autoref{fig:Dispersion}). 100 mas or 0.1\arcsec~is comparable in magnitude to the RMS error from pointing jitter as discussed earlier, and is ignored for this analysis. Furthermore, night to night variations in atmospheric dispersion due to atmospheric conditions will be subtracted out in the track specific RV offset in this analysis.
        
    \end{itemize}
    \item Far Field illumination: 
    \begin{itemize}
        \item Pupil changes - Variations in the shape and area of the effective pupil represent a change in the far-field input illumination to the fiber. We use GJ 411 as a test-bed to search for any correlation between the RVs and telescope parameters which change within the track (Section \ref{sec:analysis}).
        % \item Change in chief ray angle due to guiding errors? - Is this required?
    \end{itemize}
\end{enumerate}

Since GJ 411 is an old slowly rotating early type M dwarf, we assume that over short time periods ($\sim 30$ minutes) there is negligible astrophysical (both from the star and planet; \cite{diaz_sophie_2019, stock_carmenes_2020}) change in the stellar RVs.

\begin{figure*}
\gridline{\fig{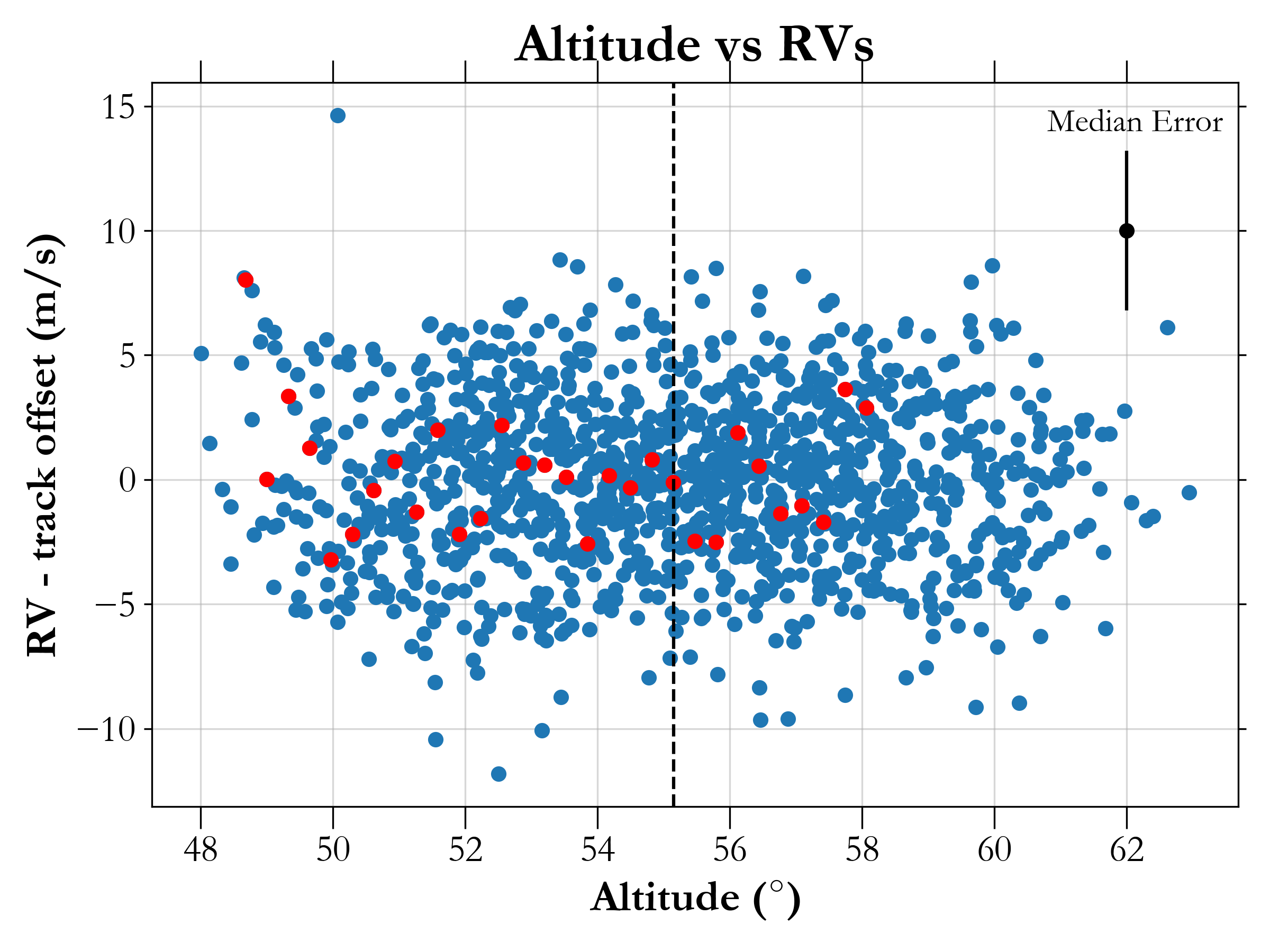}{0.42\textwidth}{{\small a) Track offset ($c_t$) subtracted RVs. There is no discernible linear trend in the data.}}    
          \fig{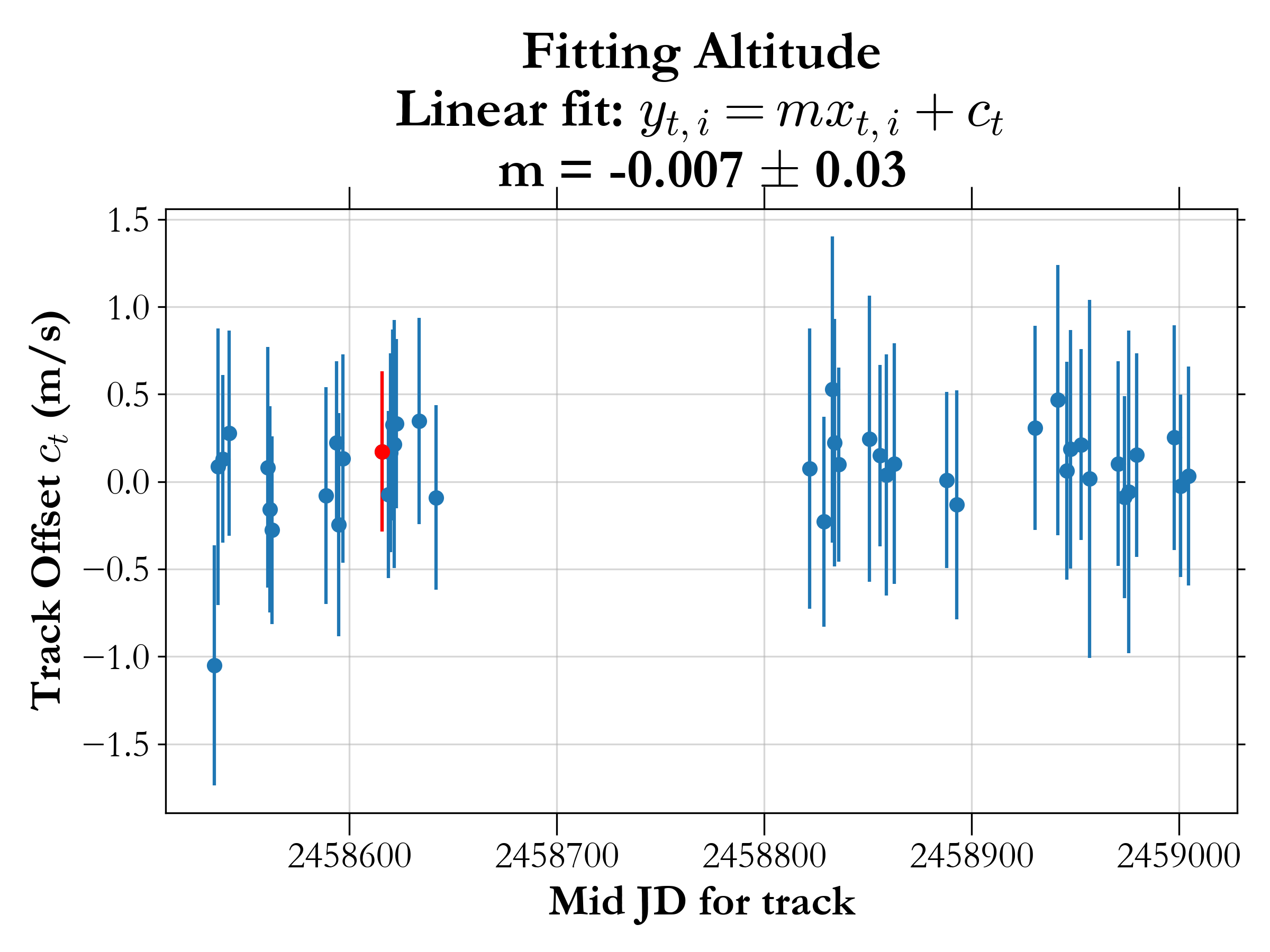}{0.45\textwidth}{ \small b) Track offset ($c_t$) as a function of time}} 
\caption{\small \textbf{a)} RVs as a function of altitude - The RVs are centered at zero since the track offset $c_t$ has been subtracted here. The median RV error is represented on the top right corner and is about 3.2 \ms{}. The vertical dashed line marks the median altitude, which is subtracted from the x axis to make both the axes zero centered. \textbf{b)} The track offset ($c_t$) that are fit, plotted as a function of time. Also mentioned is the slope obtained from the linear fit. The red points in both figures represent the track plotted in \autoref{fig:HETpupil}a, and show the scatter within a single track.}\label{fig:FitAltitude}
\end{figure*}

The ball lens scrambler images the far-field telescope illumination to the near-field of the fiber after the ball lens. This near-field is further scrambled by the output fiber bundle and is then imaged on the detector plane (Section \ref{sec:hpffiber}). If the axis corresponding to maximum change in pupil centroid (Pupil Y axis; \autoref{fig:PupilChange}a) is aligned to the dispersion axis (on HPF's detector plane), then this would have the maximum RV impact from imperfect far-field scrambling. Conversely, if it would be aligned with the cross dispersion axis then the RV impact would be minimal, and harder to measure. To place upper limits on the RV impact of pupil changes, we make the conservative assumption that changes in pupil centroids manifest along the dispersion direction. We also assume a linear relationship between pupil parameters and RVs (Section \ref{sec:lab}).

\section{Analysis and Results}\label{sec:analysis}
In order to study the impact on RVs from far-field input illumination changes due to HET's fixed altitude design, we search for correlations between the combined\footnote{Combined across East and West HET tracks} RVs and various parameters (\autoref{tab:analysis}). To obtain a robust limit we combine the RVs across different tracks by including an offset between each track\footnote{Ideally the RVs for each track should be centered at zero since we are creating a separate template for each track, which \textit{should} make this offset superfluous. However in practice this is not the case, due to variable atmospheric conditions and S/N of observations.}. This is shown below - 

\begin{equation}\label{eq:linearfit}
    y_{t, i} = m x_{t, i} + c_{t},
\end{equation}

where - 
\begin{itemize}
    \item $t$ - Index representing track number
    \item $i$ - Index representing exposures within the track (a maximum of 30)
    \item $y_{t, i}$ - observed RVs
    \item $x_{t, i}$ - Physical parameter being probed
    \item $c_t$ - Track specific offsets we fit
    \item $m$ - Slope of the fit, which is used to quantify the dependence
\end{itemize}

In practice, we include the RV errors during the fitting process, while the errors in the estimation of the pupil parameters are negligible. We also subtract the median value of $x$ ($x_0$) from $x_{t, i}$, to have both $y$ and $x$ be centered at zero. This median offset $x_0$ is degenerate with adding a constant term ($mx_0$) to the track offset $c_t$ (\autoref{eq:linearfit_mediansubtracted}).

\begin{equation}\label{eq:linearfit_mediansubtracted}
    y_{t, i} = m (x_{t, i} - x_0) + c_{t} 
\end{equation}

\subsection{Dependence of RVs on Altitude}\label{sec:fit_alt}
For our GJ 411 visits, the altitude ranges from 48$^{\circ}$ to 63$^{\circ}$ (\autoref{fig:FitAltitude}a), and is centered at the HET's fixed altitude of 55$^{\circ}$. As can be seen in \autoref{fig:PupilChange}, the peak pupil area of $\sim 50$ m$^2$ is obtained when the target is at 55$^{\circ}$, while the change in pupil area tracks the altitude well. We hence use the altitude as a proxy for change in pupil characteristics through the track.

Using \autoref{eq:linearfit_mediansubtracted}, we fit for the track offsets ($c_t$) (\autoref{fig:FitAltitude}), and the slope $m$. Combining the RVs across the tracks after subtracting a track specific offset, we obtain $m = -0.007 \pm 0.035$ \ms{} per degree change in altitude, which is consistent with the RVs being independent of changes in altitude (\autoref{fig:FitAltitude}b).

\begin{figure*}
\gridline{\fig{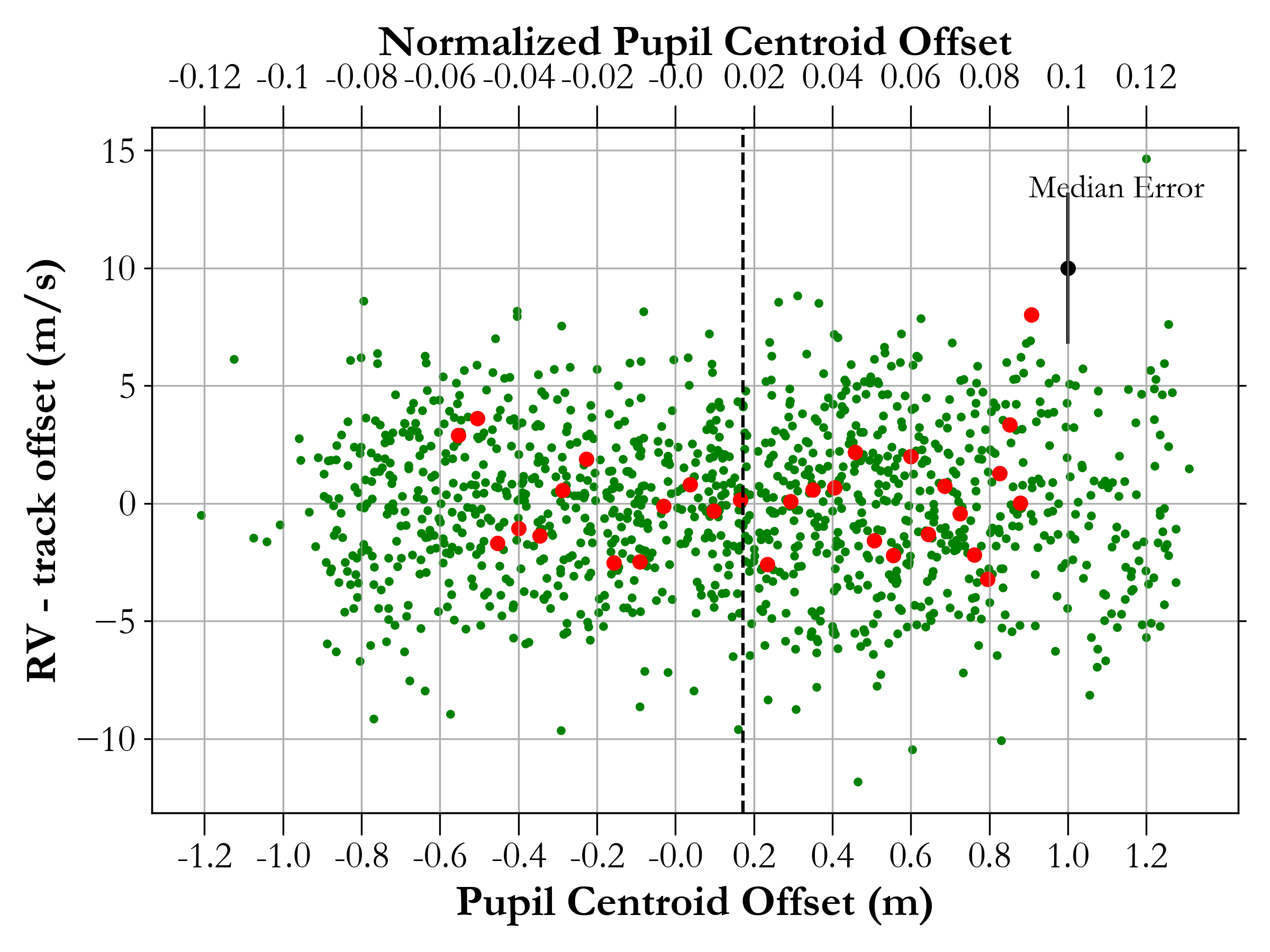}{0.42\textwidth}{{\small a) Track offset ($c_t$) subtracted RVs as a function of pupil centroid offset. There is no discernible linear trend in the data.}}    
          \fig{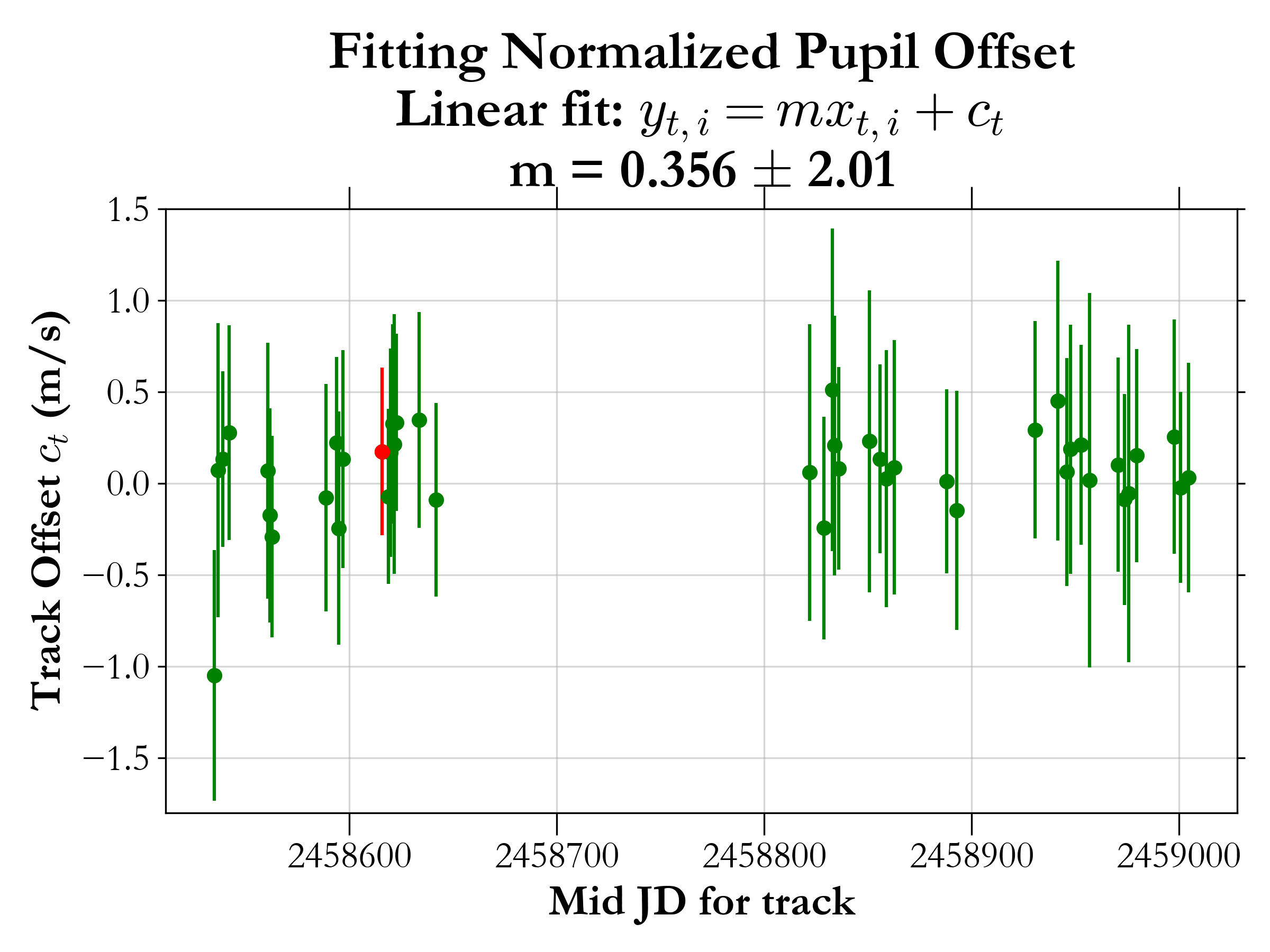}{0.45\textwidth}{ \small b) Track offset ($c_t$) as a function of time}} 
\caption{\small \textbf{a)} RVs as a function of pupil centroid offset -  The RVs are centered at zero since the track offset $c_t$ has been subtracted here. The median RV error is represented on the top right corner and is about 3.2 \ms{}. The vertical dashed line marks the median pupil offset of 0.17 m which is subtracted from the x axis to make both the axes zero centered. \textbf{b)} The track offset ($c_t$) that are fit, plotted as a function of time. Also mentioned is the slope obtained from the linear fit. The red points represent the track plotted in \autoref{fig:HETpupil}a, and show the scatter within a single track.}\label{fig:FitCentroid}
\end{figure*}

\subsection{Dependence of RVs on Pupil Centroid Offset}
We also check for dependence of the RVs on the pupil centroid. The change in pupil area is not symmetric (\autoref{fig:HETpupil}), and therefore induces a change in the centroid. We define a pupil centroid offset for each exposure, by first finding the pupil centroid, and then calculating the Euclidean distance for each exposure to the center of the mirror (0,0); we define this as pupil centroid offset. We also apply a positive or negative sign to this centroid offset based on the sign of the \textit{y} coordinate (\autoref{fig:FitCentroid}a). As can be seen in Figures \ref{fig:HETpupil} and \ref{fig:PupilChange}a, the majority of the pupil centroid shift is in the \textit{y} direction, and therefore including this sign helps distinguish between the extrema position of the pupil. Furthermore, we divide this number by the pupil diameter (10 m) to obtain the normalized pupil centroid offset. We obtain slope $m = 0.356 \pm 2.01$ \ms{} per unit change in the normalized pupil centroid  (\autoref{fig:FitCentroid}b).

% \subsection{}
As mentioned in Section \ref{sec:het}, most targets can be observed from HET in an East and West track depending on the azimuth. We separate the GJ 411 visits based on their azimuth into East and West tracks and repeat the analysis to search for correlations between RVs and altitude, and pupil centroid offset. This is done to ensure that we do not suppress a potential dependence while combining over East and West tracks.  The results from this are summarized in \autoref{tab:analysis}, and while the dependence between normalized pupil centroid offset is greater than in the combined dataset, it is still consistent with zero. We note that the uncertainties in the upper limits derived from the West track are tighter than those from the East track. This is because of two reasons: 1) The West track has 762 exposures vs 449 exposures in the East track (1.7x higher); 2) The median S/N in the West track is 480 as opposed to about 410 in the East track (1.2x higher). This is because the median seeing\footnote{Estimated as the median FWHM of the guide star in the guide camera.} in the East track is $\sim 1.50$\arcsec ~compared to $\sim 1.38$\arcsec ~(1.1x) in the West track, and that the CCAS tower obscures part of the pupil in the East track such that the median pupil area is about 41.7 m$^2$ for the East track vs 48.4 m$^2$ for the West.

\subsection{Correlation between RVs and environmental conditions}
We also perform a similar search for correlations between the RVs and other guide camera and telescope parameters such as pressure, temperature, humidity, seeing, sky brightness, and do not find a significant slope between them (\autoref{tab:analysis}).

\begin{deluxetable*}{llllllll}
\tablecaption{Slope between track offset subtracted RVs and various pupil and environmental parameters using \autoref{eq:linearfit_mediansubtracted}. \added{We also include the RVs and pupil parameters used to perform this analysis, along with the manuscript.} \label{tab:analysis}}
\tablehead{\multicolumn{4}{c}{Quantity}  & \multicolumn{3}{c}{Linear Dependence of RVs on Quantity} \\
           \colhead{Label} & \colhead{16 $\%$} & \colhead{50 $\%$} & \colhead{84 $\%$} & \colhead{$m$}   & \colhead{$\sigma_m$} & \colhead{Units for $m$}}
\startdata
\multicolumn{5}{l}{\hspace{-0.2cm}Combined (1211 exposures):}           \\
Altitude$^a$ &      51.9$^{\circ}$ &  55.2$^{\circ}$ &   58.5$^{\circ}$ &  -0.007 &    0.035 & \ms{} / degree \\
% Pupil area$^b$&     43.1 m$^2$ &            0.510 &     0.709 & \ms{} / fractional area \\
Pupil centroid offset &     -0.54 m &  0.17 m &    0.76 m &  \\
Normalized pupil offset$^b$ &     -0.054  &  0.017  &    0.076  &  0.356 &    2.01 & \ms{} / norm. pupil offset \\
\multicolumn{5}{l}{\hspace{-0.2cm}West track (762 exposures):}           \\
Altitude (W Track)$^a$ &      52.0$^{\circ}$ &  55.2$^{\circ}$ &   58.6$^{\circ}$ &          0.007 &    0.042 & \ms{} / degree \\
% Pupil area (W Track)$^b$&     43.1 m$^2$ &            -0.412 &     0.887 & \ms{} / fractional area \\
Pupil centroid offset (W Track) &    -0.62 m &  0.04 m &    0.55 m &   \\
Normalized pupil offset$^b$ (W Track) &    -0.062  &  0.004  &    0.055  &    -0.881 &    2.46 & \ms{} / norm. pupil offset \\
\multicolumn{5}{l}{\hspace{-0.2cm}East track (449 exposures):}           \\
Altitude (E Track)$^a$ &      51.6$^{\circ}$ &  55.2$^{\circ}$ &   58.4$^{\circ}$ &         -0.037 &    0.062 & \ms{} / degree \\
% Pupil area (E Track)$^b$&     43.1 m$^2$ &            2.148 &     1.182 & \ms{} / fractional area \\
Pupil centroid offset (E Track) &     -0.16 m &  0.49 m &    1.05 m &    \\
Normalized pupil offset$^b$ (E Track) &     -0.016  &  0.049  &    0.105  &    2.88 &    3.51 & \ms{} / norm. pupil offset \\
\multicolumn{5}{l}{\hspace{-0.2cm}Environmental conditions:}           \\
Ambient Temp$^a$ &      4.9$^{\circ}$C & 14.2$^{\circ}$C & 18.5$^{\circ}$C  &  -0.174 &    0.499 & \ms{} / $^{\circ}$C \\
Humidity$^a$ &          17.7~$\%$ &      33.4~$\%$ &     54.8~$\%$ &     0.043 &     0.074 & \ms{} / $\%$ \\
Pressure$^a$ &          798 torr &  802 torr & 805 torr &     1.224 &         0.717 & \ms{} / torr \\
Seeing (FWHM)$^a$ &     1.23\arcsec & 1.42\arcsec & 1.87\arcsec &          0.164 &         0.523 & \ms{} / \arcsec \\
Sky Brightness$^a$  &   17.8 &   19.0 &      20.0 &         -0.016 &        0.082 & \ms{} / mag \\
\enddata
\tablenotetext{a}{Fit performed on median subtracted quantity}
% \tablenotetext{b}{Fractional change with respect to median}
\tablenotetext{b}{Normalized by pupil diameter (10 m)}
% \tablenotetext{c}{Normalized by pupil diameter (10 m)}
\end{deluxetable*}

\section{Conclusion and Summary}\label{sec:conclusion}
\subsection{Far-field scrambling results for HPF}\label{sec:conclusion_hpf}

We present a detailed study of the sensitivity of high precision RV measurements to telescope illumination variations, leveraging on-sky data from the HPF instrument at HET. The HET pupil illumination systematically varies across the full span of GJ 411 RVs, providing a harsh but important test of the fiber delivery system and spectrometer. We explore a variety of possible correlations between illumination offsets and the recorded RVs, and rule out sensitivities for the GJ 411 RVs at the  $4.6\pm 26$ \cms{} level\footnote{\textit{$|$m$|$} $\times$  ($\Delta$ centroid offset) = 0.356 $\times$ (0.076 - (-0.054)) ~$\simeq 4.6$ \cms{}} for the combined GJ 411 RVs, which is consistent with our expectations from lab tests of 18 $\pm$ 1 \cms{}. When we separate the RVs by HET track based on azimuth, we see a worst case dependence of RVs with normalized pupil centroid offset of $35\pm 42$ \cms{} for the East track\footnote{We note that this worst case scenario is driven by East track observations of GJ 411 where a portion of the pupil is blocked by the CCAS tower.}.

In Section \ref{sec:analysis} we fit a linear relationship between the HPF RVs for GJ 411 and various pupil and environmental parameters. To make this fit more robust we combine the RVs across the entire observing run, and subtract track specific offsets (\autoref{eq:linearfit_mediansubtracted}), to account for near-field effects, atmospheric conditions, and other systematics which would affect this analysis. Doing so enables us to combine the entire RV dataset of 1211 points and gauge the effectiveness of the HPF far-field scrambling system (Section \ref{sec:hpf}). We show that the RVs are independent of the drastic changes in pupil parameters as the target altitude changes (\autoref{tab:analysis}), and the pupil centroid shifts (\autoref{fig:PupilChange}a). Therefore we validate the far-field scrambling performance of the ball lens system developed by \citep{halverson_efficient_2015} for HPF.

\subsection{Importance for other EPRV instruments}
HPF is a NIR instrument in search of habitable zone planets around M dwarfs; since these planets have RV semi-amplitudes greater than Earth analogues, HPF's instrumental precision goal is not 10 \cms{}.  However the next generation of precision RV instruments in search of an Earth analogue with a Doppler signal of 10 \cms{} have instrumental precision goals aiming sub-10 \cms{}. At these precision levels, despite being illuminated by telescopes with more conventional pupil designs, these pupil effects matter. This harsh test of far-field scrambling with HPF, allows us to probe in a macroscopic manner the subtle effects that would affect future precision RV instruments.

 The ball lens double scrambler design presents an efficient and compact solution which offers high scrambling gain. We demonstrate a harsh on-sky test which validates the scrambling performance of this system, and demonstrates its performance for future instruments. For conventional telescopes we do not expect pupil centroid offsets at the macroscopic level seen in HET ($\sim 10 \%$). However even if there were 1$\%$ offsets, our current best estimate\footnote{Estimated from the dependence of the RVs on the  normalized pupil offset across all observations.} for the performance of similar scrambling system is about $0.46 \pm 2.6$ \cms{}, while the worst case scenario\footnote{Estimated from the dependence of the RVs on the normalized pupil offset, as observed during the East track at HET.} would be about $3.5 \pm 4.2$ \cms{}.  NEID \citep{halverson_comprehensive_2016, schwab_design_2016} has a fiber train similar to that used in HPF, including its double scrambler. The limits placed on the on-sky performance of this system confirms that it can meet the performance needs for NEID, and other similar instruments. We demonstrate that the far-field scrambling related error can be limited to a very small fraction of the 10 \cms{} RV error budget (added in quadrature).

While conventional telescopes do not see such extreme pupil variations, for the next generation of large telescopes with multi-segmented mirrors, there may be variations in the pupil from night to night due to difference in reflectivity between segments. At the same time, there can be holes in the mirror due to segments being replaced. These pupil changes would cause spurious RV offsets if the fiber input far-field illumination is not scrambled well enough. We therefore validate the on-sky performance of a scrambling system which helps reduce the RV errors due to changes in the far-field illumination pattern, as well as demonstrate a test which can be used for other instruments with varying input illumination patterns. 

% \footnote{One could also replicate such a test using an artificial pupil mask with a stable illumination using a light source with high information content}.

\section*{Acknowledgement}
 % CEHW 
This work was partially supported by funding from the Center for Exoplanets and Habitable Worlds. The Center for Exoplanets and Habitable Worlds is supported by the Pennsylvania State University, the Eberly College of Science, and the Pennsylvania Space Grant Consortium. These results are based on observations obtained with the Habitable-zone Planet Finder Spectrograph on the HET. We acknowledge support from NSF grants AST-1006676, AST-1126413, AST-1310885, AST-1517592, AST-1310875, AST-1910954, AST-1907622, AST-1909506, ATI 2009889, ATI 2009982, and the NASA Astrobiology Institute (NNA09DA76A) in our pursuit of precision radial velocities in the NIR. We acknowledge support from the Heising-Simons Foundation via grant 2017-0494. Computations for this research were performed on the Pennsylvania State University’s Institute for Computational and Data Sciences’ Roar supercomputer,including the CyberLAMP cluster supported by NSF grant MRI-1626251.

The Hobby-Eberly Telescope is a joint project of the University of Texas at Austin, the Pennsylvania State University, Ludwig-Maximilians-Universität München, and Georg-August Universität Gottingen. The HET is named in honor of its principal benefactors, William P. Hobby and Robert E. Eberly. The HET collaboration acknowledges the support and resources from the Texas Advanced Computing Center. We thank the HET staff for continued support and their dedication to the facility and for the skillful execution of our observations of our observations with HPF.
%ADS
This research has made use of NASA's Astrophysics Data System Bibliographic Services. 
% JPL (for Sam)
Part of this research was carried out at the Jet Propulsion Laboratory, California Institute of Technology, under a contract with the National Aeronautics and Space Administration (NASA). SK would like to acknowledge Walter Eugene O'Reilly and Theodora for help with this project.

\facilities{ HET (HPF)}
\software{
\texttt{astroquery} \citep{ginsburg_astroquery_2013}, 
\texttt{astropy} \citep{robitaille_astropy:_2013, price-whelan_astropy_2018},
\texttt{barycorrpy} \citep{kanodia_python_2018}, 
\texttt{HxRGproc} \citep{ninan_habitable-zone_2018},
\texttt{ipython} \citep{perez_ipython:_2007}
\texttt{matplotlib} \citep{hunter_matplotlib:_2007},
\texttt{mc3} \citep{cubillos_correlated-noise_2017}
\texttt{numpy} \citep{oliphant_numpy:_2006},
\texttt{pandas} \citep{mckinney_data_2010},
\texttt{pyHETobs}, (This work)
\texttt{scipy} \citep{oliphant_python_2007, virtanen_scipy_2020},
\texttt{SERVAL} \citep{zechmeister_spectrum_2018}
}

\appendix 

\section{pyHETobs}\label{sec:pyHETobs}
In order to estimate the size and shape of the HET pupil as a function of time, we developed a \texttt{Python} package titled \texttt{pyHETobs}\footnote{\href{https://indiajoe.github.io/pyHETobs/}{https://indiajoe.github.io/pyHETobs/}}. It calculates the geometry of the moving pupil on the segmented spherical primary mirror, taking into account the obscuration caused by the WFC, support structures, as well as the CCAS tower (in the East track). The parked azimuth of the telescope, and the stellar coordinate in sky completely defines the telescope's effective pupil. The tool has features to drop custom segments in the primary mirror which were not available for the night from the pupil calculation. It also enables users to plan observations to maximise the pupil area during the exposures. In the documentation we have included example files to calculate the pupil centroid, area and make plots similar to the ones in this manuscript. 

\bibliography{MyLibrary}

\listofchanges\end{document}